\documentclass[10pt,pre,aps,twocolumn,superscriptaddress,floatfix,notitlepage]{revtex4-2}

\usepackage[latin1]{inputenc}
\usepackage{amsmath,amssymb,amsfonts,bm}
\usepackage{mathrsfs,dsfont}
\usepackage{graphicx}
\usepackage{braket}
\usepackage{natbib}
\usepackage[dvipsnames]{xcolor}
\usepackage[normalem]{ulem}

\usepackage{ragged2e}

\usepackage{hyperref}
\hypersetup{
    colorlinks  = true,
    citecolor   = blue,
    linkcolor   = blue,
    urlcolor    = blue,
    filecolor   = magenta,  
    pdftitle    = {},
    pdfpagemode = FullScreen,
    }

\usepackage{comment}

\begin{document}
\title{The Transient Counting Statistics of Autonomous Quantum Clocks}
\author{Oscar Arandes}
\email{oscar.arandes@fysik.su.se}
\affiliation{Department of Physics, Stockholm University, AlbaNova University Center, 10691 Stockholm, Sweden}

\author{Sreenath K. Manikandan}
\email{skm@tifrh.res.in}
\affiliation{Tata Institute of Fundamental Research Hyderabad, 36/P, Gopanpally Village, Serilingampally Mandal, Hyderabad, Telangana 500046, India}

\begin{abstract}
 At the heart of every atomic clock is a quantum coherent emitter---the laser---that could in principle be treated autonomously. While their frequency or phase stability is typically studied in the semiclassical regime of a large number of emitted quanta, recent works on autonomous quantum clocks suggest that the few-quanta regime can offer statistical information in the temporal domain. This provides complementary means to probe the fundamental limits to precise timekeeping in the quantum regime. Motivated by these works, we consider simple models for two- and three-level quantum emitters subject to an athermal fueling through dephasing, which generically results in coherences in the steady state. A large deviation principle with finite-time contributions is used to identify the transient counting statistics and their dependence on the initial conditions. We conclude by discussing the implications for precise timekeeping and quantum sensing using autonomous quantum emitters as clocks in their transient regime.
\end{abstract}
\date{\today}

\maketitle

\section{Introduction}
Simple models of atomic clocks consist of a laser that acts as a quantum coherent emitter of a time-continuous optical field. This field serves as the frequency source, feedback stabilized by an atom acting as the frequency standard~\cite{AtomClockReview}. The driving laser is the real clock. From a thermodynamic perspective, a laser is also one of the earliest studied models for a quantum engine~\cite{LaserEngine} that is autonomous (consuming elementary thermodynamic resources), motivating a thermodynamic description for the working principles of a clock. It is therefore natural to expect that, like any other device, the clockwork of an atomic clock must be fundamentally constrained by the laws of quantum statistical physics and thermodynamics~\cite{erker_autonomous_2017,milburn_thermodynamics_2020}, applied to the laser that produces the heartbeat of every atomic clock. Being inherently quantum mechanical, the clock is also, in principle, affected by the very observations made on it, influencing its accuracy and precision~\cite{Manikandan2023,he_measurement_2022,Benny_Manikandan}.

Characterizing these with modern atomic clocks is, however, a challenging task. Clock standardization measurements primarily probe the frequency or phase stability of the oscillating field~\cite{vanier_quantum_2024,AtomClockReview,schulte_prospects_2020,jiang_making_2011,Chou}. This typically makes use of measurements in the semiclassical regime, where the average number of photons from the emitter is $\langle N\rangle_t \gg 1$. With frequency stabilization based on radiative transitions between atomic levels, highly precise atomic clocks are now in operation. The best available optical atomic lattice clocks drift by approximately one second over the age of the universe~\cite{zheng_differential_2022,bloom_optical_2014}. Given that these timescales are so tiny, one is motivated to ask whether the quantum statistical features that determine the ultimate quantum bounds on the accuracy and precision of an atomic clock are observable at all.

Questions of this nature are of fundamental interest. One possible approach to such clock characterization experiments is to use a newer version of the atomic clock, with improved resolution, to benchmark an older one, i.e., to benchmark a slower clock against a faster clock reference. We consider a closely related approach, using simple models of lasers as few-level quantum systems to provide the clock reference. Recent works probing the fundamental thermodynamic limits to precise timekeeping with such systems suggest that, in principle, the quantum thermodynamic constraints we are interested in could be observable in short-time experiments in the temporal domain~\cite{erker_autonomous_2017,Meier2023,MeierMinoguchi2025,Manikandan2023,Benny_Manikandan,schwarzhans_autonomous_2021,pearson_measuring_2021,AutonomousReview2024,he_measurement_2022,milburn_thermodynamics_2020,singh_quantum_2026}. Such short-time experiments give access to the few-quanta regime ($\langle N\rangle_t \sim 1$) of the counting statistics of clock ticks, where one could treat the clock as a small system in quantum thermodynamics. This offers a perspective in the temporal domain that is complementary to the phase or frequency standardization measurements in the semiclassical limit~\cite{vanier_quantum_2024,AtomClockReview,schulte_prospects_2020,jiang_making_2011,Chou}.
 
To have quantum-controllable few-level systems, we assume the use of radiative transitions in an artificial atom to implement the clock transitions. In such a setting, the autonomy of the clock is ensured by considering quantum emitters fueled by elementary quantum thermodynamic resources~\cite{AutonomousReview2024,erker_autonomous_2017}. In particular, we use dephasing quantum noise as the fuel~\cite{Manikandan2023,Benny_Manikandan}, modeled using Hermitian dissipators that do not commute with the emitter's Hamiltonian. A good artificial atomic clock for the quantum-clock-characterization studies we are interested in would be a rapidly decaying superconducting qubit with a short lifetime, typically of microseconds, driven by dephasing quantum noise. Such scenarios have gained considerable interest in recent years in quantum thermodynamics, as refrigerators~\cite{revealingFuel,Sundelin2026} and as autonomous quantum clocks~\cite{Manikandan2023,Benny_Manikandan,AutonomousReview2024}.

\begin{figure*}[htbp]
    \centering
    \begin{minipage}{0.40\textwidth}
        \centering
        \includegraphics[width=\linewidth]{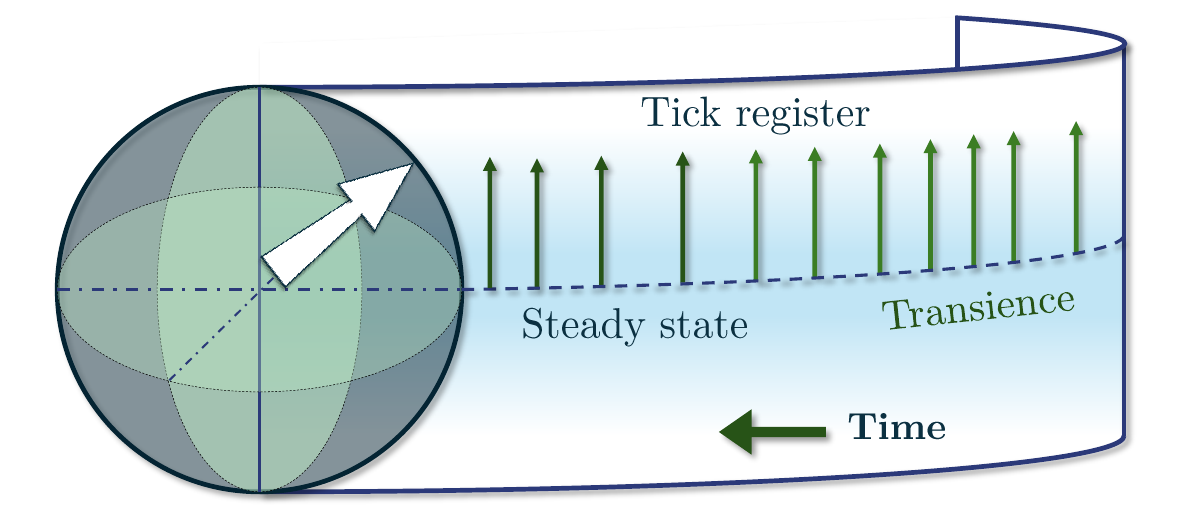}
    \end{minipage}
    \hspace{0.5cm}  
    \begin{minipage}{0.45\textwidth}
        \centering
        \includegraphics[width=\linewidth]{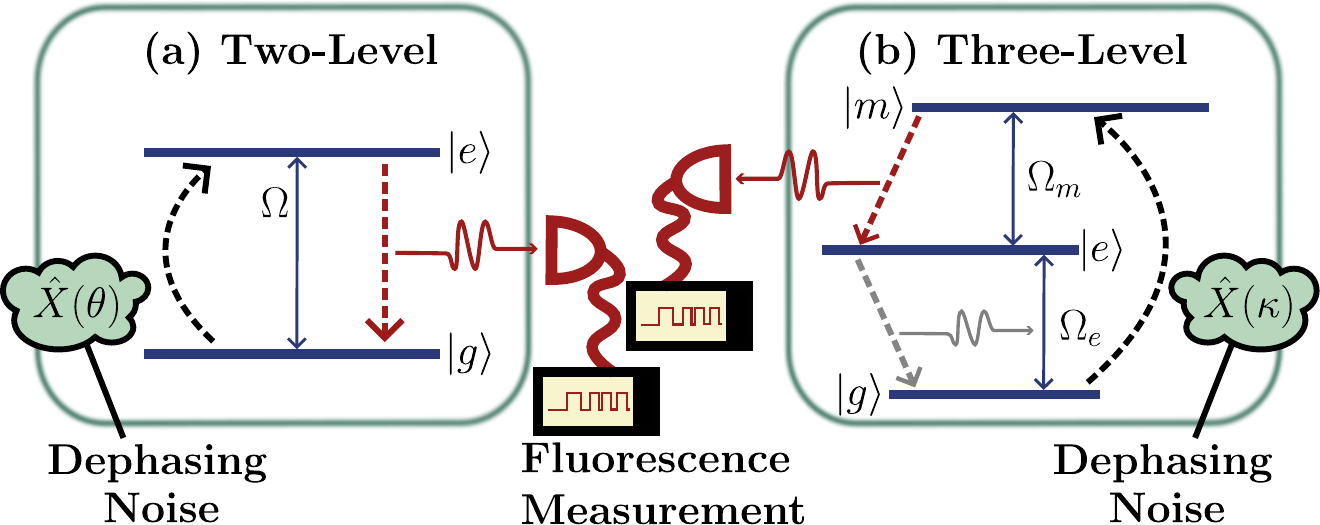}
    \end{minipage}
    \caption{ \justifying   
    Schematic of the clock dynamics. Left: An initial qubit state (Bloch sphere) drives a tick register, where fluorescence events are recorded as ticks along the time axis. Reading from right to left, the dynamics evolve from the initial state through the transient regime toward the steady state.
    Right: A quantum emitter is subject to two processes: dephasing noise generated by the operator $\hat{X}$ (green) and  photodetection of the emitted fluorescence (red), whose discrete clicks constitute the ticks of the clock. The dephasing drives the population and closes the emission cycle, while the counted fluorescence yields the counting statistics analyzed in the text. (a)~Two-level emitter: the counted decay $|e\rangle\!\to\!|g\rangle$ at rate $\gamma_w$ (red) is detected, and the dephasing on $\hat{X}$ re-pumps the population. (b)~Three-level cascade: the counted transition $|m\rangle\!\to\!|e\rangle$ at rate $\gamma_w$ (red) is detected, while the subsequent decay $|e\rangle\!\to\!|g\rangle$ at rate $\gamma_e$ (grey) is lost to the reservoir and produces no signal; the dephasing re-pumps $|g\rangle\!\to\!|m\rangle$, closing the cascade.
    }
    \label{fig:setup}
\end{figure*}

While earlier works have motivated such scenarios for autonomous timekeeping devices~\cite{Manikandan2023,Benny_Manikandan,erker_autonomous_2017}, they were restricted to the large-but-finite-time regime and to steady states without coherence. The transient response and the role of initial conditions, where one expects the genuine quantum effects of the emitter to show up, remained elusive. Our present work overcomes these limitations by considering scenarios that generically lead to coherence in the steady state of the emitter. To account for the quantum effects in the emission characteristics relevant both in the transient regime and in the steady state, we present a revised large deviation principle with transient contributions that depend on the initial conditions of the quantum emitter. In this regard, our work goes beyond earlier works that have characterized the large-but-finite-time behavior within the large-deviation approach~\cite{LecomteAppertRolland2007, GarrahanJackLecomte2007, GarrahanJackLecomte2009, GarrahanLesanovsky2010, AtesOlmosGarrahan2012, Touchette2009, Touchette2018, CarolloGarrahanLesanovsky2018, CarolloGarrahan2019, Perfetto2022, Manikandan2023, LandiReview2024,MenczelFlindtBrange2026,Benny_Manikandan}, which provide the leading-order contribution to the statistics we compute.  Expressions valid at finite observation times do appear in Refs.~\onlinecite{GarrahanJackLecomte2009, CarolloGarrahanLesanovsky2018}, and the dependence on the initial state that survives at long times is noted in Ref.~\onlinecite{LandiReview2024}. There, however, these features are considered primarily to establish that they do not affect the asymptotic rates. Our aim is instead to retain and characterize these transient initial-state contributions. Although our objectives differ, the machinery developed in the following sections is closely related to that used in these works. The limitations and the need to go beyond leading-order estimates were also noted in Ref.~\onlinecite{Benny_Manikandan}. 

Our models, when extended to account for external drives, also relate to quantum sensing strategies~\cite{Qsensing,LandiReview2024,bilinskaya_quantum_2026,Klaus1,Klaus2} for drives in the transient and steady state of the clock, which is indeed an important practical application of the statistics we compute. We discuss an exemplary scenario in greater detail in Appendix~\ref{app:driven}.  The setups we describe are very much within reach of experiments~\cite{entropyCost,he_measurement_2022,pearson_measuring_2021}, given that they can be simulated with remarkable quantum measurement and control efficiencies, for instance using superconducting quantum circuits. 

We reemphasize that, while such artificial quantum emitters may have interesting timekeeping applications, they need not be the best choices to function as timekeeping devices, given that better quantum clocks are available. Nevertheless, our analysis reinforces the idea that such artificial atom-based clocks could serve as quantum systems with remarkable quantum control capabilities, which can be used to study the fundamental limitations of precise timekeeping and quantum sensing in the transient and steady-state regimes. 

This article is organized as follows. We first revisit the large-deviation approach to the counting statistics of quantum emitters subject to Markovian dynamics in Sec.~\ref{sec:TheoreticalMethods}. In Sec.~\ref{sec:Quantum_emitters}, we then consider two- and three-level autonomous quantum emitters fueled by dephasing noise as examples, and provide a systematic characterization of the statistics in both the transient and steady-state regimes. To support the theoretical results, we perform exact numerical simulations using measurement (Kraus) operators (see Appendices~\ref{app:Kraus_Formalism}--\ref{app:Kraus_Continuous}). We conclude in Sec.~\ref{sec:figuresofmerit} by discussing the implications of our results for precise timekeeping and quantum sensing using autonomous quantum clocks.

\section{Theoretical methods} \label{sec:TheoreticalMethods}
   
We now introduce a framework based on \textit{large deviation theory (LDT)} for Markovian open quantum systems~\cite{GarrahanLesanovsky2010,Perfetto2022}, to characterize the time-domain counting statistics of quantum jump events in a coherent quantum emitter fueled by dephasing quantum noise. Our treatment includes both transient and steady-state-coherence contributions that go beyond standard scenarios that lead to coherence-less steady states. The Markovian dynamics of the quantum state $\hat{\rho}$ of the emitter will generically be described by the Lindblad master equation $\frac{d}{dt}\hat{\rho}(t) = \hat{\mathcal{L}}[\hat{\rho}(t)]$, where the Liouvillian $\hat{\mathcal{L}}$ is given by~\cite{breuer_theory_2007}
\begin{equation}
 \hat{\mathcal{L}}[\hat{\rho}] = -i[\hat{H},\hat{\rho}] + \sum_{\mu=1}^{N_L} \mathcal{D}[\hat{L}_\mu]\hat{\rho}  , 
\end{equation}
with 
${\mathcal{D}[\hat{L}_\mu]\hat{\rho} =
   \hat{L}_\mu \hat{\rho} \hat{L}_\mu^\dagger
   - \frac{1}{2}\{\hat{L}_\mu^\dagger \hat{L}_\mu,\hat{\rho}\} } $
being the dissipator. Here, $\hat{L}_\mu$ denotes the $\mu$-th of the $N_L$ Lindblad operators describing the dissipative processes, where we have absorbed the rates into the Lindblad operators for simplicity. The dephasing noise fueling the quantum emitter in our case corresponds to the scenario where one of the Lindblad operators, $\hat{L}_\mu$, is Hermitian.

The dynamics can be unraveled into individual quantum trajectories, each characterized by the number of jumps that have occurred in the observed quantum measurement channel, which, in our case, corresponds to measuring the clock's ticking. To this end, let $\hat{\rho}(N,t)$ be the \textit{conditional} density operator describing the occurrence of $N$ jumps~\cite{ZollerMarteWalls1987}. The total density operator is then recovered by summing over all possible values of $N$:
\begin{equation}
    \hat{\rho}(t) = \sum_{N=0}^{\infty} \hat{\rho}(N,t).
\end{equation}
The probability $\mathcal{P}_t(N)$ of observing $N$ jumps in the time interval $(0,t)$ is given by,
\begin{equation}
    \mathcal{P}_t(N) = \mathrm{tr}\!\left[\hat{\rho}(N,t)\right],
    \label{eq:Pt(N)}
\end{equation}
and the normalization condition reads,
\begin{equation}
    \sum_{N=0}^{\infty} \mathcal{P}_t(N) = \sum_{N=0}^{\infty} \mathrm{tr}\!\left[\hat{\rho}(N,t)\right] = 1.
\end{equation}
In the context of Lindblad dynamics, $\mathcal{P}_t(N)$ acquires a large-deviation form at long times~\cite{GarrahanLesanovsky2010, Touchette2018}. The corresponding moment generating function (MGF) can be expressed as,
\begin{equation}
    Z_t(s) = \mathrm{tr}\! \left[ \sum_{N=0}^{\infty} \hat{\rho}(N,t) \, e^{-sN} \right]
    \equiv \mathrm{tr}\!\left[ \hat{\rho}_s(t) \right],
    \label{eq:MGF_trace}
\end{equation}
where one defines the \textit{tilted} density operator $\hat{\rho}_s(t)$ as the discrete Laplace transform of $\hat{\rho}(N,t)$:
\begin{equation}
    \hat{\rho}_s(t) = \sum_{N=0}^{\infty} \hat{\rho}(N,t)\, e^{-sN}.
\end{equation}
Starting from the set of equations obeyed by $\hat{\rho}(N,t)$~\cite{ZollerMarteWalls1987}, and applying the Laplace transformation introduced above, one finds that the tilted density operator evolves according to a generalized master equation 
${\frac{d}{dt}\hat{\rho}_s(t) = \hat{\mathcal{L}}_s[\hat{\rho}_s(t)]}$~\cite{ZhengBrown2003,Brown2006Review,EspositoHarbolaMukamel2009}, where the tilted Liouvillian is given by
\begin{equation}
    \begin{split}
    \hat{\mathcal{L}}_s [\hat{\rho}_s] &= 
    -i[\hat{H}, \hat{\rho}_s]
    + \sum_{\mu \neq w}\mathcal{D}[\hat{L}_\mu]\hat{\rho}_s \\
    &+ e^{-s}\, \hat{L}_w \hat{\rho}_s \hat{L}_w^\dagger
    - \frac{1}{2}\{ \hat{L}_w^\dagger \hat{L}_w,\hat{\rho}_s \}.
    \end{split}
\end{equation}
The sum above runs over all decay channels $\{ \hat{L}_\mu\}$ except $\hat{L}_w$, which denotes the particular decay channel whose quantum jumps are being counted (as ticks of the clock in our examples). The tilting parameter, or counting field, $s$ thus biases the jump statistics associated with the channel $\hat{L}_w$. Note that, unlike the original Liouvillian generator, $\hat{\mathcal{L}}_s$ is \emph{not} trace-preserving. Therefore, $\hat{\rho}_s(t)$ does not represent a physical density operator but instead acts as a generating object whose trace gives the MGF of the number of jumps (c.f. Eq.~\eqref{eq:MGF_trace}).

We defer to upcoming sections involving examples regarding the details of how the counting statistics are inferred from simulations of two- and three-level quantum emitters fueled by dephasing quantum noise, which naturally account for all transient and steady state contributions. To obtain the counting statistics analytically that account for the transient and steady-state emission characteristics using the tilted Lindbladian, we employ a vectorized representation of the density matrix, $\hat{\rho}(t)\rightarrow |\rho(t)\rangle\!\rangle$. The tilted-Lindblad evolution equation then becomes
\begin{equation}
  \frac{d}{dt} |\rho(t)\rangle\!\rangle = \mathcal{L}_s |\rho(t)\rangle\!\rangle,
\end{equation}
where $\mathcal{L}_s$ is a $d^2 \times d^2$ matrix representation of the tilted Liouvillian, with $d$ the dimension of the Hilbert space. The formal solution is given by
$| \rho(t) \rangle \! \rangle = e^{t \mathcal{L}_s} \, | \rho(0) \rangle \! \rangle$.
Further details on the vectorization procedure are provided in Appendix~\ref{app:vectorization}. The right and left eigenvectors of $\mathcal{L}_s$ in Liouville space, associated with eigenvalues $\lambda_k$, are denoted by $| r_k\rangle \! \rangle$ and $\langle \! \langle l_k |$, respectively. The dependence on the parameter $s$ is explicitly stated. They satisfy
\begin{equation}
   \mathcal{L}_s \, | r_k^{(s)} \rangle \! \rangle = \lambda_k \, | r_k^{(s)} \rangle \! \rangle, \quad 
   \langle \! \langle l_k^{(s)} | \, \mathcal{L}_s = \lambda_k \, \langle \! \langle l_k^{(s)} |,
\end{equation}
with biorthogonal normalization $\langle \! \langle l_k^{(s)} | r_{k'}^{(s)} \rangle \! \rangle = \delta_{k,k'}$.
Assuming that the Liouvillian is diagonalizable, we can expand the propagator acting on the vectorized initial density matrix $| \rho(0) \rangle \! \rangle$ in the spectral basis of $\mathcal{L}_s$:
\begin{equation}
   e^{t\mathcal{L}_s} | \rho (0) \rangle \! \rangle
   = \sum_k e^{t \lambda_k(s)} \, | r_k^{(s)} \rangle \! \rangle \, \langle \! \langle l_k^{(s)} | \rho (0) \rangle \! \rangle .
\end{equation}
Consequently, the MGF $Z_t(s)$ can be naturally expressed within the Liouvillian formalism as (see Appendix~\ref{app:vectorization})
\begin{equation}
    Z_t(s) = \sum_k c_k(s) \, e^{t \lambda_k(s)},
    \label{eq:MGF_vectorized}
\end{equation}
where we have defined
\begin{equation}
    c_k(s) \equiv \langle \! \langle \mathds{1} | r_k^{(s)} \rangle \! \rangle \, \langle \! \langle l_k^{(s)} | \rho (0) \rangle \! \rangle.
    \label{eq:ck}
\end{equation}
The eigenvectors are defined only up to a normalization: for any nonzero function $f_k(s)$, the rescaling
$ |r_k^{(s)}\rangle\!\rangle \to f_k(s)\,|r_k^{(s)}\rangle\!\rangle, 
  \langle\!\langle l_k^{(s)}| \to f_k^{-1}(s)\,\langle\!\langle l_k^{(s)}| $
leaves the eigenvalue problem and the biorthonormality condition invariant. The coefficients $c_k(s)$ in Eq.~\eqref{eq:ck} are constructed to be invariant under this rescaling. Consequently, $Z_t(s)$ and every cumulant derived from it are independent of the choice of normalization, as required. We exploit this freedom to fix a convenient gauge for the dominant mode, imposing $\langle\!\langle \mathds{1} | r_{\mathrm{dom}}^{(s)} \rangle\!\rangle = 1$ for all $s$. 

In this gauge, the dominant coefficient reduces to
\begin{equation}
c_{\mathrm{dom}}(s) = \langle\!\langle l_{\mathrm{dom}}^{(s)} | \rho(0) \rangle\!\rangle ,
\end{equation}
making explicit that the entire dependence of the prefactor on the initial state is carried by the dominant left eigenmode.

Assuming that the leading eigenvalue $\lambda_{\rm dom}(s)$ is non-degenerate, separated from the rest of the spectrum by a finite gap, and has non-zero overlap $c_{\rm dom}(s) \neq 0$, we can write the MGF as a spectral sum,
\begin{equation}
  Z_t(s) = e^{t \lambda_{\mathrm{dom}}(s)}
  \Bigg(
  c_{\mathrm{dom}}(s)
   + \sum_{k \neq \mathrm{dom}} c_k(s)\,
   e^{-t \Delta_k(s)}
   \Bigg),\label{specsum}
\end{equation}
where $c_{\mathrm{dom}}$ denotes the coefficient in Eq.~\eqref{eq:ck}
corresponding to the leading eigenvalue and we have defined ${\Delta_k(s) = \lambda_{\mathrm{dom}}(s) - \lambda_k(s)}$. Factoring out the dominant contribution and defining the spectral gap
$\Delta(s) \equiv \min_{k \neq \mathrm{dom}} \mathrm{Re}\,\Delta_k(s)$,
each subleading contribution is suppressed by at least $e^{-t\Delta(s)}$, so that
\begin{equation}
    Z_t(s) = e^{t \lambda_{\mathrm{dom}}(s)}\,c_{\mathrm{dom}}(s)
    \left[1 + \mathcal{O}\!\left(e^{-t \Delta(s)}\right)\right].
    \label{eq:Zt(s)}
\end{equation}
Taking the logarithm, and using $\ln(1+x) = \mathcal{O}(x)$ for $x \to 0$
with $x = \mathcal{O}(e^{-t\Delta})$, we obtain (also see~\cite{LandiReview2024})
\begin{equation}
    \ln Z_t(s) = t\,\lambda_{\mathrm{dom}}(s) + \ln c_{\mathrm{dom}}(s)
   + \mathcal{O}\!\left(e^{-t \Delta(s)}\right).
\end{equation}
The average number of jumps is then given by  
\begin{equation}
    \langle N \rangle_t
    = -\lambda'_{\mathrm{dom}}(0)\, t
    - (\ln c_{\mathrm{dom}})'(0) 
    + \mathcal{O}\!\left(e^{-t\Delta(0)}\right) .
\end{equation}
Here and in the following, the notation $\lambda'_{\rm dom}(0)$ denotes derivatives with respect to $s$ evaluated at $s=0$, i.e., $\lambda'_{\rm dom}(0) = \partial_s\lambda_{\rm dom}(s)|_{s=0}$. By keeping only the leading correction from the initial-state overlap, we then arrive at
\begin{equation}
  \langle N \rangle_t \simeq  k_{\rm ss} t - (\ln c_{\mathrm{dom}})'(0)  ,
  \label{eq:mean_with_correc}
\end{equation}
where $k_{\rm ss} = - \lambda'_{\rm dom}(0)$, and $\simeq$ denotes equality up to terms that vanish exponentially in $t$. Similarly, the variance is given by
\begin{equation}
   \langle N^2 \rangle_t - \langle N \rangle_t^2  \simeq  v_{\rm ss} t + (\ln c_{\mathrm{dom}})''(0)  ,
  \label{eq:var_with_correc} 
\end{equation}
with $v_{\rm ss} \!=\! \lambda''_{\rm dom}(0)$ and
$(\ln c_{\mathrm{dom}})''(0) \!=\! c''_{\rm dom}(0)/c_{\rm dom}(0) - \left( c'_{\rm dom}(0)/c_{\rm dom}(0) \right)^{\!2} $. Equations~\eqref{eq:mean_with_correc}--\eqref{eq:var_with_correc} constitute the central focus of this work. As we show in the following sections, they are to be considered to accurately capture the emission statistics of autonomous quantum coherent emitters as clocks across different regimes of interest here.

The derivation above has revealed that the transient effects are encoded in the left eigenvector of the tilted Liouvillian, through the prefactor $c_{\rm dom}(s)$. While conceptually transparent, evaluating $c_{\rm dom}(s)$ directly is rarely the most practical route, as the dependence of the eigenvectors on $s$ becomes analytically intractable beyond the simplest cases. Crucially, however, every quantity in Eq.~\eqref{eq:mean_with_correc} is evaluated at $s=0$, where the dominant eigenpair is known exactly $\lambda_{\rm dom}(0)=0$, $\langle\!\langle l_{\rm dom}^{(0)}|=\langle\!\langle\mathds{1}|$, $|r_{\rm dom}^{(0)}\rangle\!\rangle=|\rho_{\rm ss}\rangle\!\rangle$. This invites a perturbative treatment in $s$ around the steady state~\cite{flindt_full_2004,FlindtNovotnyBraggio2010}, which both renders the transient contributions computable in closed form and clarifies their physical origin. It can be shown that Eq.~\eqref{eq:mean_with_correc} can be recast as (see Appendix~\ref{app:pert_expansion})
\begin{equation}
   \langle N\rangle_t
   \simeq
   k_{\rm ss}t
   + \int_0^\infty d\tau\,
   \left[k(\tau)-k_{\rm ss}\right].
\end{equation}
Here, $k(t) = \gamma_w\, \mathrm{tr}\!\left[
\hat{L}_w^\dagger \hat{L}_w \hat{\rho}(t)\right]$, $\hat{\rho}(t) = e^{t\hat{\mathcal{L}}_0}\hat{\rho}(0)$, is the instantaneous photon-emission rate of the untilted state at time $t$, and $k_{\rm ss} = k(\infty) = -\lambda'_{\rm dom}(0) $ is its stationary value. The physical meaning becomes then manifest: the leading, time-extensive term counts emissions at the stationary rate, while the correction is the integrated excess (or deficit) of emissions accumulated during the transient relaxation from the initial state $| \rho(0) \rangle\!\rangle$ to the steady state $| \rho_{\rm ss} \rangle\!\rangle$. In particular, the correction vanishes identically when $| \rho(0) \rangle\!\rangle=| \rho_{\rm ss} \rangle\!\rangle$, as it must, since the system then emits at the stationary rate at all times.

A parallel interpretation holds for the variance. The time-extensive term $\lambda_{\rm dom}''(0)\,t$ is the stationary variance of the emission process. The correction $(\ln c_{\rm dom})''(0)$ consists of two physically distinct contributions (see Appendix~\ref{app:pert_expansion}). The first depends on the initial state and, analogous to its counterpart in the mean, encodes the excess fluctuations accumulated while $| \rho(0) \rangle\!\rangle$ relaxes towards the steady state $| \rho_{\rm ss} \rangle\!\rangle$. Consequently, it vanishes when $| \rho(0) \rangle\!\rangle=| \rho_{\rm ss} \rangle\!\rangle$.  The second contribution is independent of the initial state and remains present even when the system starts in the steady state, arising from emissions near the boundaries of the time observation window whose correlations are only partially captured. This boundary effect gives a constant contribution that does not grow with $t$, in contrast to the extensive term.

\section{Quantum emitters with coherence in their steady state}
\label{sec:Quantum_emitters}

To make use of the formalism presented above to characterize simple models of autonomous quantum clocks, let us first consider a simple two-level transition driven by dephasing quantum noise. The guided resonant emission between the two-levels is observed continuously in time to serve as the clock signal. Later, we also discuss a three-level example for completeness, which allows us to account for losses that are unmonitored. The photodetection efficiency, however, is assumed to be unity in all the examples considered; hence, we do not account for any dark counts, but our models and the statistics can be further generalized to account for these additional sources of imperfections as well.

\subsection{Two-level system}

The Hamiltonian of the two-level quantum emitter is chosen to be ($\hbar = 1$)
\begin{equation}
    \hat{H} = \Omega \ket{e}\bra{e} \,,
\end{equation}
where $\ket{e}$ denotes the excited state of the two-level system, and we assume that the frequency $\Omega$ is experimentally accessible through resonance fluorescence measurement with an ideal photodetector. To fuel the emitter using minimal thermodynamic resources, we assume that the two-level system undergoes dephasing along the axis defined by the Hermitian observable $\hat{X}$, described by the master equation $\frac{d \hat{\rho}}{dt} = -i [\hat{H}, \hat{\rho}] - \gamma_m [\hat{X},[\hat{X}, \hat{\rho}]]$. Equivalently, the same dynamics can be engineered via continuous weak measurement of $\hat{X}$ on the two-level system at measurement rate $\gamma_m$~\cite{Manikandan2023,Benny_Manikandan}, provided that the measurement record is discarded such that the resulting evolution is unconditional. We consider a one-parameter generalization of the choice $\hat{X}=\hat{\sigma}_x$ used in Refs.~\cite{Manikandan2023,Benny_Manikandan},
\begin{equation}
    \hat{X}(\theta) = \hat{\sigma}_x \cos\theta + \hat{\sigma}_z \sin\theta .
    \label{eq:X_2level}
\end{equation}
The fluorescence-monitored qubit is governed by the interplay between the energy basis $\{|g\rangle,|e\rangle\}$ and the noise basis defined by the eigenstates of $\hat X(\theta)$. The angle $\theta$ parametrizes this misalignment, ranging from aligned bases at $\theta=\pi/2$ ($\hat X=\hat\sigma_z$) to maximal misalignment at $\theta=0$ ($\hat X=\hat\sigma_x$). This mismatch fully determines the steady state, including the emergence of energy-basis coherences.

The clock signal is obtained from a detector counting fluorescence emission between the two levels, where the detector integration time $dt$ is much smaller than $\gamma_w^{-1}$, $\gamma_w$ being the spontaneous emission rate. In other words, we assume that the integration time is much shorter than the characteristic decay time from the excited to the ground state (see Fig.~\ref{fig:setup} for a schematic of the fluorescence and dephasing setup).

The tilted Liouvillian of the two-level system relevant for obtaining the counting statistics is given by 
\begin{equation}
    \begin{split}
    \hat{\mathcal{L}}_s[\hat{\rho}] &= -i [\hat{H}, \hat{\rho}] -\gamma_m [\hat{X}, [\hat{X},\hat{\rho}]] \\
    &+ \gamma_w \left[  e^{-s} \, \hat{\sigma}_- \hat{\rho} \, \hat{\sigma}_+ - \frac{1}{2} \left( \hat{\rho} \, \hat{\sigma}_+  \hat{\sigma}_- +  \hat{\sigma}_+ \hat{\sigma}_- \, \hat{\rho}  \right) \right] ,
   \end{split}
   \label{eq:lindblad_twolevel}
\end{equation}
where $\hat{L}_w = \hat{\sigma}_- = \ket{g}\bra{e}$ and $\hat{L}_w^\dagger =\hat{\sigma}_+ = \ket{e}\bra{g}$. 

To understand the dynamics let us set the tilt $s=0$ and write the density matrix in terms of its Bloch components,
$\hat{\rho} = \frac{1}{2}\left(\hat{\mathds{1}}+x\hat{\sigma}_x+y\hat{\sigma}_y+z\hat{\sigma}_z\right)$,
with $z=p_e-p_g$, so that Eq.~\eqref{eq:lindblad_twolevel} becomes
\begin{align}
    \dot{x} &= - \Omega  y
    -\left(4\gamma_m\sin^2\theta+\frac{\gamma_w}{2}\right)x
    +2\gamma_m\sin(2\theta)\,z, \label{eq:Bloch_x} \\
    \dot{y} &=  \Omega  x-\zeta \, y, \label{eq:Bloch_y} \\
    \dot{z} &= 2\gamma_m \sin(2\theta)\,x- \left[4\gamma_m \cos^2\theta+\gamma_w\right] z-\gamma_w ,
    \label{eq:Bloch_z}
\end{align}
with $\zeta = \frac{1}{2}\gamma_w + 4\gamma_m$. Let us first build an intuition for the dynamics. Ignoring the coherences, Eq.~\eqref{eq:Bloch_z} shows that the dephasing operator $\hat{X}$ induces the population dynamics,
\begin{equation}
    \dot{p}_e
    =\epsilon(\theta)\,(p_g-p_e) -\gamma_w p_e, \quad
    \epsilon(\theta)=2\gamma_m\cos^2\theta,
    \label{eq:pee_simpfld}
\end{equation}
where $\epsilon(\theta)$ is the noise-induced transition rate.
Note that it is the $\hat{\sigma}_x$ component of $\hat{X}$ that drives
transitions between $|g\rangle$ and $|e\rangle$; the noise is the only mechanism able to populate the excited state, since the Hamiltonian is diagonal in the energy basis. In this simplified picture, the rate equation in Eq.~\eqref{eq:pee_simpfld} yields the steady-state population $p_e^{\rm ss}\to \frac{\epsilon(\theta)}{2\epsilon(\theta)+\gamma_w}$.
 
However, the same noise process that induces population transitions also creates coherences. The crucial term is the source $2\gamma_m\sin(2\theta)\,z$ in Eq.~\eqref{eq:Bloch_x}: a noise kick acting on a state with population imbalance ($z\neq0$) generates a coherence (except at $\theta=\{0,\frac{\pi}{2}\}$). The dephasing in turn damps this coherence, both directly, through the rate $4\gamma_m\sin^2\theta+\frac{\gamma_w}{2}$ acting
on the $x$ quadrature, and indirectly, through the Hamiltonian precession, which rotates $x$ into the orthogonal quadrature $y$ where it is damped at the rate $\zeta$, opening an additional channel for coherence decay. What makes this scenario non-trivial is that these coherences are not merely a by-product of the dynamics: they feed back on the populations, entering through the term $2\gamma_m\sin(2\theta)\,x$ in Eq.~\eqref{eq:Bloch_z}. 

Solving the stationary Bloch equations exactly, we obtain the steady state,
\begin{equation}
  \rho_{\rm ss} =
  \begin{pmatrix}
    p_e^{\rm ss} & \rho_u - i\rho_v \\
    \rho_u + i\rho_v & 1 - p_e^{\rm ss}
  \end{pmatrix} ,
  \label{eq:2level_density_ss}
\end{equation}
where the excited-state population is
\begin{align}
  p_e^{\rm ss} &= \frac{\epsilon_{\rm eff}(\theta)}
  {2\,\epsilon_{\rm eff}(\theta) + \gamma_w} , \qquad
  \epsilon_{\rm eff}(\theta) = \epsilon(\theta)\,R(\theta),
  \label{eq:pe_ss}
\end{align}
with
\begin{equation}
  R(\theta) = \frac{\tfrac{1}{2}\gamma_w + \frac{\Omega^2}{\zeta}}
  {4\gamma_m\sin^2\theta + \tfrac{1}{2}\gamma_w + \frac{\Omega^2}{\zeta}} .
\end{equation}
The coherence feedback renormalizes the population dynamics with the effective rate $\epsilon_{\rm eff}(\theta)$. The suppression factor $R(\theta)\in(0,1]$ thus captures the net effect of the coherence feedback on the stationary population. The simple rate-equation picture ($R\to1$) is recovered in two limits: at $\theta=0$,
where no coherence is generated, and for $\Omega^2\gg\gamma_m\zeta$, where rapid precession drains the coherence before it can feed back. 

Finally, the steady-state coherences are
\begin{align}
  \rho_u = \frac{\chi(\theta)}{2}\,\big(2p_e^{\rm ss} - 1\big),  \qquad
  \rho_v = \frac{\Omega}{\zeta}\,\rho_u ,
\end{align}
where
\begin{equation}
   \chi(\theta) = \frac{2\gamma_m\sin(2\theta)}
  {4\gamma_m\sin^2\theta + \tfrac{1}{2}\gamma_w + \frac{\Omega^2}{\zeta}} .
\end{equation}
This makes the origin of the off-diagonal elements in
Eq.~\eqref{eq:2level_density_ss} explicit: the coherence $x=\chi(\theta)\,z$ is nonzero whenever the population is imbalanced ($z\neq0)$ and the noise axis is tilted ($\sin(2\theta)\neq0$).

\begin{figure}[htbp]
   \includegraphics[width=0.45\textwidth]{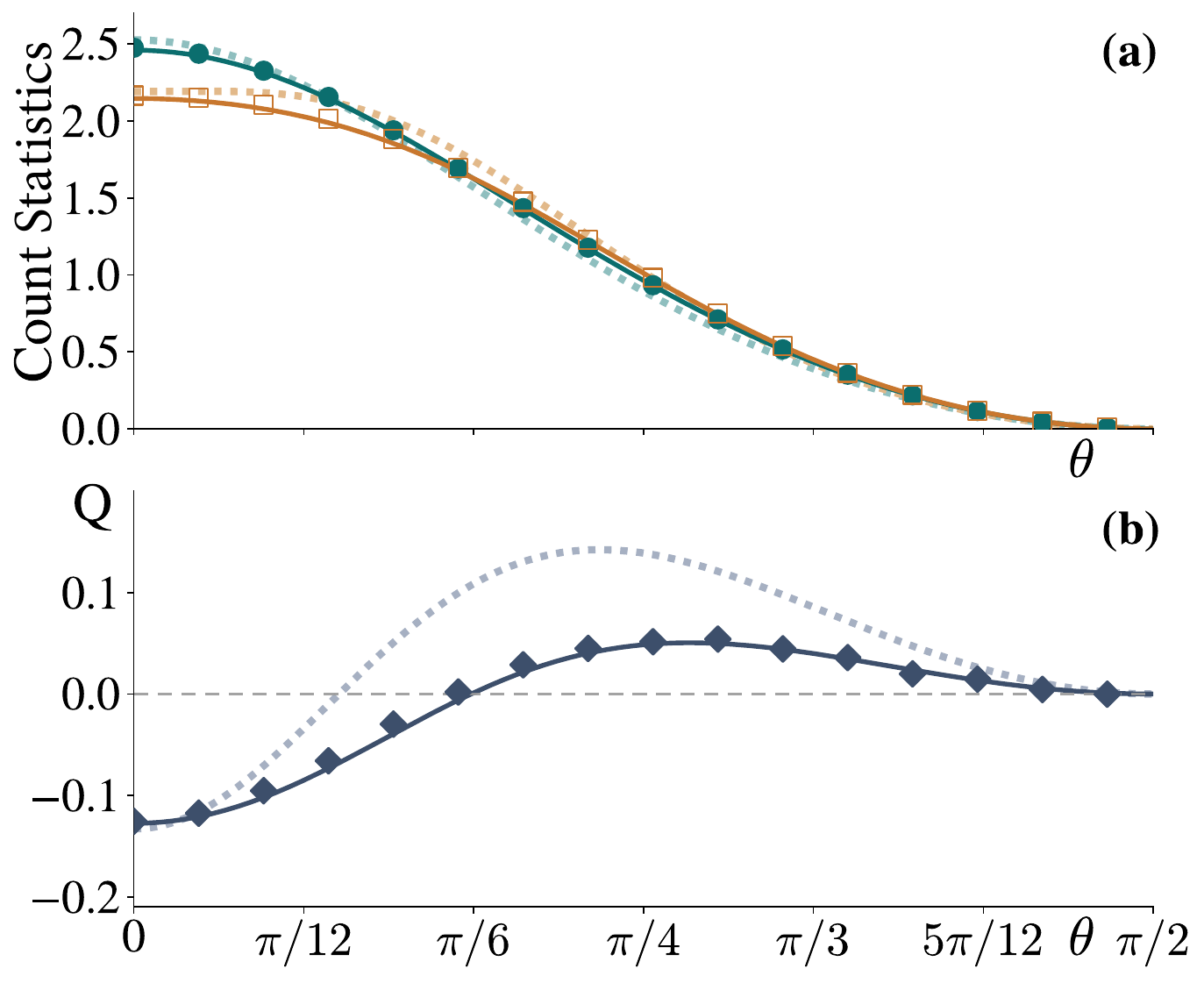}
   \caption{
    (a) Mean number (green) and variance (orange) of clicks as functions of $\theta$ for the observable $\hat{X}(\theta)$. 
    (b) Mandel $Q_t$ parameter as a function of $\theta$ (dark blue). 
    In both panels, dashed lines denote the LDT predictions, while solid lines include both the LDT and transient contributions arising from the initial state (Eqs.~\eqref{eq:app_mean_summary} and~\eqref{eq:app_var_summary} evaluated analytically). Scatter points correspond to numerical simulations performed using the Kraus-operator formalism (see Appendix~\ref{app:Kraus_Formalism}). Parameters: $\Omega = 1$, $\gamma_w = 6$, and $\gamma_m = 8$. Simulations were performed with time step $dt = 0.001$ over $1000$ steps using $10^6$ trajectories. Initial state $\hat{\rho}(0) = \ket{g}\bra{g}$. }
   \label{fig:2level_mean_variance_Q_ground}
\end{figure}

\begin{figure}[htbp]
   \includegraphics[width=0.45\textwidth]{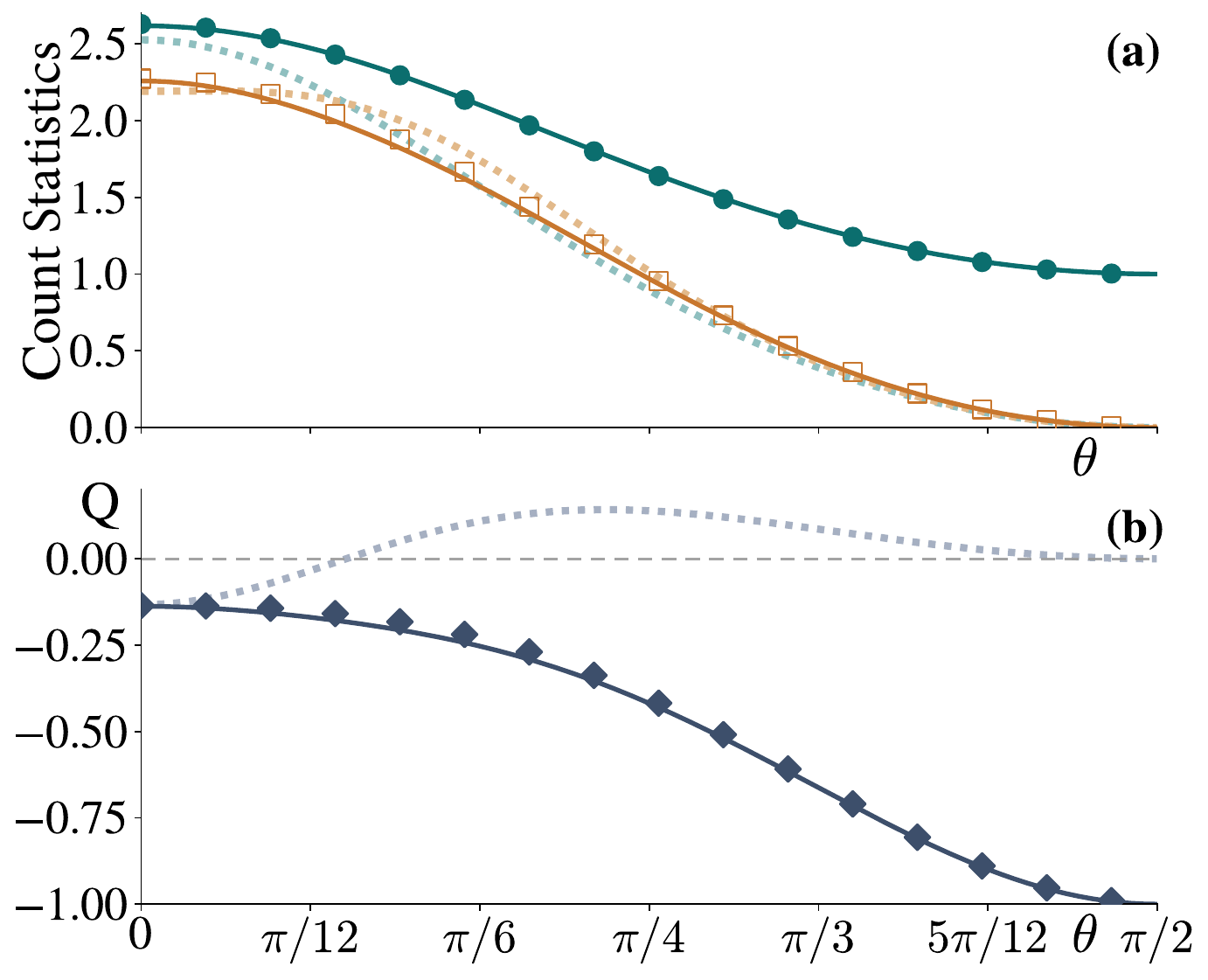}
   \caption{
    (a) Mean number (green) and variance (orange) of clicks as functions of $\theta$ for the observable $\hat{X}(\theta)$. 
    (b) Mandel $Q_t$ parameter as a function of $\theta$ (dark blue). 
    In both panels, dashed lines denote the LDT predictions, while solid lines include both the LDT and transient contributions arising from the initial state (Eqs.~\eqref{eq:app_mean_summary} and~\eqref{eq:app_var_summary} evaluated analytically). Scatter points correspond to numerical simulations performed using the Kraus-operator formalism (see Appendix~\ref{app:Kraus_Formalism}). Parameters: $\Omega = 1$, $\gamma_w = 6$, and $\gamma_m = 8$. Simulations were performed with time step $dt = 0.001$ over $1000$ steps using $10^6$ trajectories. 
    Initial state $\hat{\rho}(0) = \ket{e}\bra{e}$. }
   \label{fig:2level_mean_variance_Q_excited}
\end{figure}

To characterize the statistics of the quantum jumps we now consider the tilted dynamics with $s\neq 0$. We evaluate the mean number of clicks $\langle N\rangle_t$ and their variance $\mathrm{Var}(N)_t = \langle N^2\rangle_t - \langle N\rangle_t^2$ as functions of $\theta$. The tick statistics are further characterized by the Mandel $Q_t$ parameter~\cite{Mandel1995}, $Q_t = \frac{\langle N^2\rangle_t - \langle N\rangle_t^2}{\langle N\rangle_t} - 1 $, which quantifies deviations from Poissonian statistics ($Q_t=0$). At finite $t$, it depends strongly on the initial state, with a marked difference between preparing the system in the ground state (Fig.~\ref{fig:2level_mean_variance_Q_ground}) and in the excited state (Fig.~\ref{fig:2level_mean_variance_Q_excited}). In both cases, the leading time-extensive term predicted by LDT fails to reproduce the exact simulations of quantum jumps using the operator sum (Kraus) representation. Agreement is only recovered by including the transient contributions.

As observed in Figs.~\ref{fig:2level_mean_variance_Q_ground} and~\ref{fig:2level_mean_variance_Q_excited}, the leading-order LDT prediction yields a vanishing mean emission rate as $\theta\to\pi/2$. This follows from the stationary emission rate, $k_{\rm ss}(\theta)=\gamma_w p_e^{\rm ss}(\theta)$. Although $R(\theta)$ remains finite as $\theta\to\pi/2$, the effective pump scales as $\epsilon_{\rm eff}\propto\cos^2\theta$. The emission rate is therefore increasingly pump-limited as $\theta$ approaches $\pi/2$, and vanishes exactly at $\theta=\pi/2$ where the noise-induced pumping is switched off. Once the qubit decays it can never be re-excited.

From the perspective of the tilted Liouvillian, this behavior is reflected in the fact that, at $\theta=\pi/2$, the spectrum becomes independent of $s$. In particular, the dominant eigenvalue satisfies $\lambda_{\rm dom}(s)=0$ for \textit{all} $s$, implying that all time-extensive cumulant rates vanish and the LDT carries no information. All $s$-dependence of the counting statistics is instead encoded in the prefactor $c_{\rm dom}(s)= p_e(0)\,e^{-s} + p_g(0)$, i.e., in the eigenvectors of $\hat{\mathcal{L}}_s$. Consequently, the outcome depends crucially on the initial-state preparation. 

Near $\theta=\pi/2$, the steady state approaches the ground state. A system prepared in $|g\rangle$ therefore starts arbitrarily close to the steady state, and since the ground state produces no emission, both the mean count, $\langle N\rangle_\infty \to 0$, and transient contributions vanish (since $c_{\rm dom}(s) \to 1$). In this case, the leading-order and corrected predictions around $\theta=\pi/2$ coincide (Fig.~\ref{fig:2level_mean_variance_Q_ground}). 

By contrast, a system prepared in $|e\rangle$ is initially maximally far from the steady state. At $\theta=\pi/2$, it emits exactly one photon before falling dark, giving $\langle N\rangle_\infty=1$ given sufficient time, as shown in Fig.~\ref{fig:2level_mean_variance_Q_excited}. Note that maximally sub-Poissonian statistics ($Q_t \rightarrow -1$) here cannot be equated with maximal accuracy, since it means a near-deterministic count (one click) given sufficient time, rather than regular ticks. We return to this in Subsection~\ref{sec:timekeeping}.

\begin{figure}[htbp]
   \includegraphics[width=0.45\textwidth]{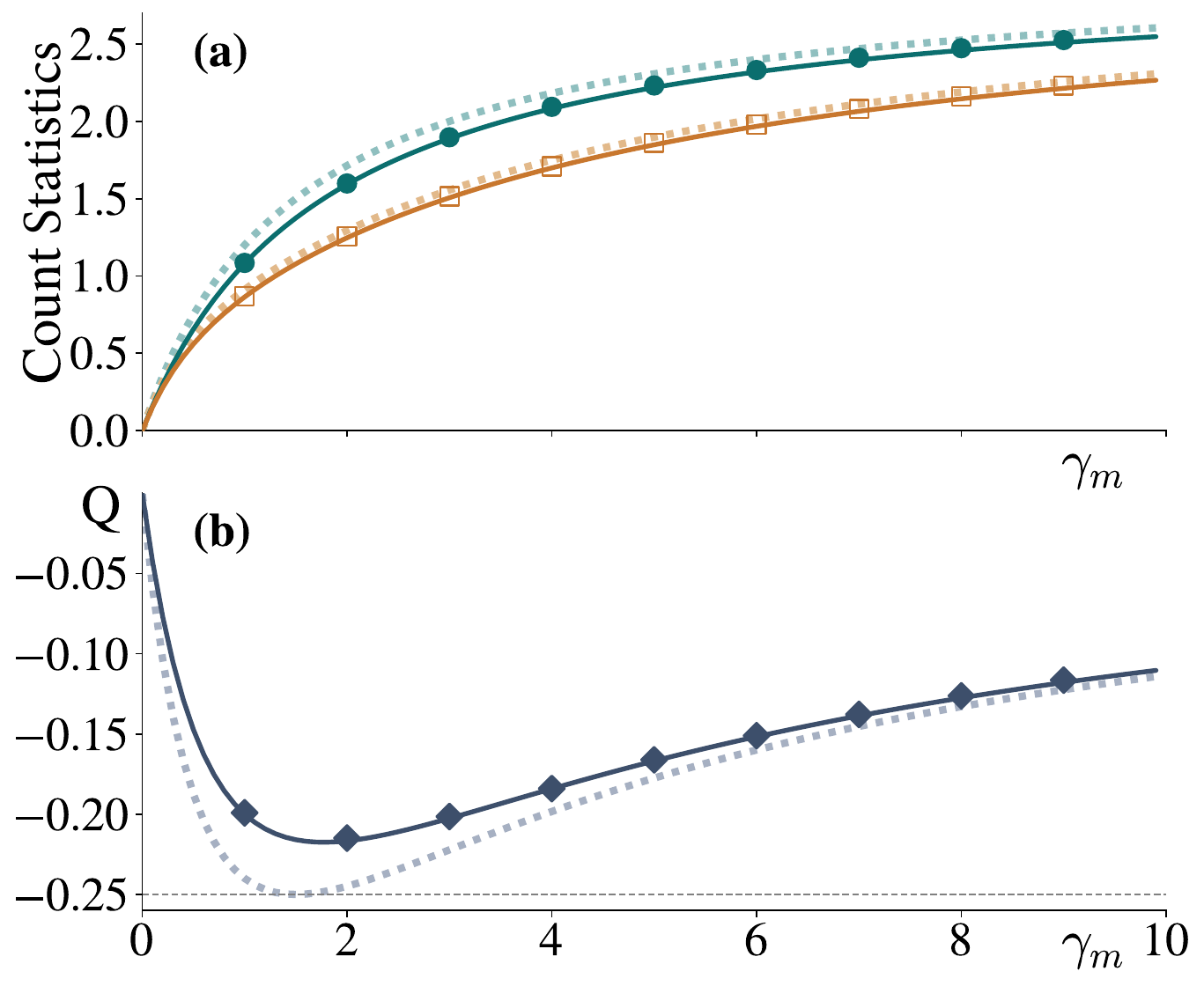}
   \caption{
    (a) Mean number (green) and variance (orange) of clicks as functions of $\gamma_m$ for the observable $\hat{X}(\theta=0) = \hat{\sigma}_x$. 
    (b) Mandel $Q_t$ parameter as a function of $\gamma_m$ (dark blue). 
    In both panels, dashed lines denote the leading-order predictions (see Eqs.~\eqref{eq:mean_sigma_x_ground_t},  \eqref{eq:variance_sigma_x_ground_t}), while solid lines include both the LDT and transient contributions arising from the initial state (see Eqs.~\eqref{eq:mean_sigma_x_ground_early}, \eqref{eq:variance_sigma_x_ground_early} and \eqref{eq:Q_sigma_x_ground_early}). Scatter points correspond to numerical simulations performed using the Kraus-operator formalism (see Appendix~\ref{app:Kraus_Formalism}). Parameters: $\Omega = 1$, $\gamma_w = 6$. The simulations are performed with time step $dt = 0.001$ over $1000$ steps using $10^6$ trajectories. Initial state $\hat{\rho}(0) = \ket{g}\bra{g}$. 
    }
   \label{fig:2level_mean_variance_ground_gm}
\end{figure}

\begin{figure}[htbp]
   \includegraphics[width=0.45\textwidth]{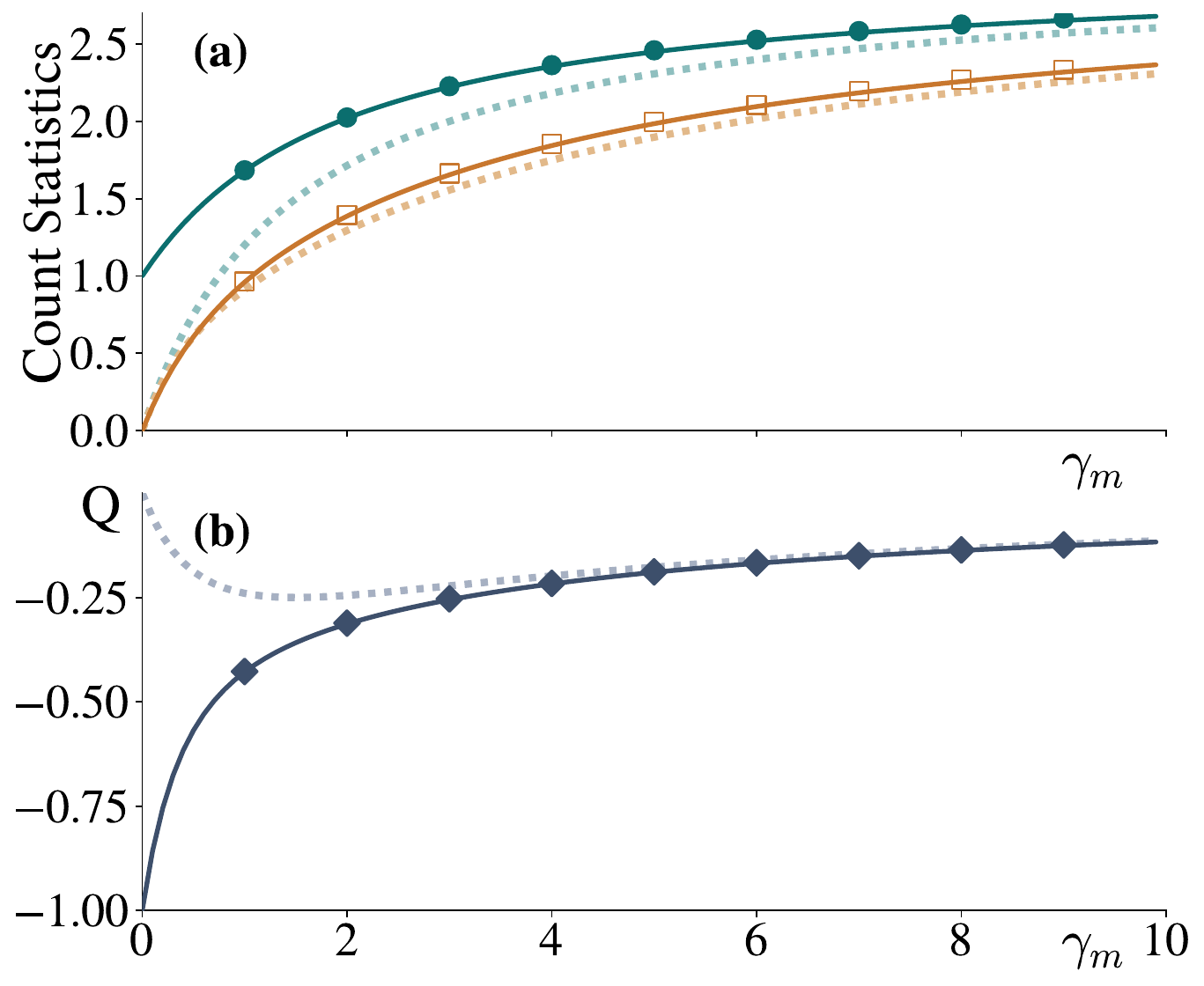}
   \caption{
    (a) Mean number (green) and variance (orange) of clicks as functions of $\gamma_m$ for the observable $\hat{X}(\theta=0) = \hat{\sigma}_x$. 
    (b) Mandel $Q_t$ parameter as a function of $\gamma_m$ (dark blue).
    In both panels, dashed lines denote the LDT predictions, while solid lines include both the LDT and transient contributions arising from the initial state (see Eqs.~\eqref{eq:mean_sigma_x_excited_early}, \eqref{eq:variance_sigma_x_excited_early} and \eqref{eq:Q_sigma_x_excited_early}). Scatter points correspond to numerical simulations performed using the Kraus-operator formalism (see Appendix~\ref{app:Kraus_Formalism}). Parameters: $\Omega = 1$, $\gamma_w = 6$. The simulations are performed with time step $dt = 0.001$ over $1000$ steps using $10^6$ trajectories. Initial state $\hat{\rho}(0) = \ket{e}\bra{e}$. 
    }
   \label{fig:2level_mean_variance_excited_gm}
\end{figure}

We can alternatively fix the angle, for instance to ${\theta=0}$, which both yields simple analytical expressions (there are no coherences) and lets us examine the behaviour as the monitoring strength $\gamma_m$ is varied. For the ground state as the initial state, the average number of clicks $\langle N \rangle_t = {\langle N \rangle_t}_{\mathrm{LDT}} + {\langle N \rangle}_{\mathrm{early}}^{(g)}$ is
\begin{align}
    {\langle N \rangle_t}_{\mathrm{LDT}} &=  \frac{2\gamma_m\gamma_w}{4\gamma_m+\gamma_w}t, \label{eq:mean_sigma_x_ground_t} \\[1ex]
     {\langle N \rangle}_{\mathrm{early}}^{(g)} &= - \frac{2\gamma_m \gamma_w}{(4\gamma_m+\gamma_w)^2} , \label{eq:mean_sigma_x_ground_early}   
\end{align}
where the second term is the transient contribution from the initial state. Similarly, the variance
$\mathrm{Var}(N)_t
    = \mathrm{Var}_{\mathrm{LDT}}(N)_t + \mathrm{Var}_{\mathrm{early}}^{(g)}(N)$
reads
\begin{align}
    \mathrm{Var}_{\mathrm{LDT}}(N)_t &= \frac{2\gamma_m \gamma_w (16\gamma_m^2 + 4\gamma_m\gamma_w+\gamma_w^2)}{(4\gamma_m+\gamma_w)^3}t ,\label{eq:variance_sigma_x_ground_t} \\[1ex]
    \mathrm{Var}_{\mathrm{early}}^{(g)}(N) &= -\frac{2\gamma_m \gamma_w(16\gamma_m^2 -2 \gamma_m \gamma_w+\gamma_w^2)}{(4\gamma_m+\gamma_w)^4} ,
    \label{eq:variance_sigma_x_ground_early}
\end{align}
and the Mandel $Q^{(g)}_t$ parameter is 
\begin{equation}
    Q^{(g)}_t = -\frac{2 \gamma_m \gamma_w }{(4 \gamma_m + \gamma_w)^2 } 
    \frac{2 (4\gamma_m + \gamma_w)t -5}{(4 \gamma_m + \gamma_w)t -1}.
    \label{eq:Q_sigma_x_ground_early}
\end{equation}
These quantities are shown as functions of $\gamma_m$ in Fig.~\ref{fig:2level_mean_variance_ground_gm}. Here the initial-state dependence is \emph{erased} at strong dephasing. The ground- and excited-state predictions (discussed below) differ at weak dephasing but converge as $\gamma_m$ grows, both approaching the leading-order result. A large $\gamma_m$ makes the pumping and relaxation fast, so the system reaches its steady state almost immediately and the transient contributions are over before an appreciable number of photons has been counted. The system forgets its preparation quickly, and the transient contributions (which measure the distance of the initial state from stationarity) become negligible relative to the time-extensive term. 
For an excited-state initial condition (Fig.~\ref{fig:2level_mean_variance_excited_gm}), the deviation from the leading-order prediction is much stronger than for the ground state. The average number of clicks is $\langle N \rangle_t = {\langle N \rangle_t}_{\mathrm{LDT}} + {\langle N \rangle}_{\mathrm{early}}^{(e)}$, with the same time-extensive term~\eqref{eq:mean_sigma_x_ground_t} and
\begin{equation}
     {\langle N \rangle}_{\mathrm{early}}^{(e)} =  \frac{\gamma_w (2\gamma_m +\gamma_w)}{(4\gamma_m+\gamma_w)^2} .
    \label{eq:mean_sigma_x_excited_early}
\end{equation}
Note that this correction is \emph{positive}, in contrast to the ground-state case: an excited qubit emits more during its transient than a stationary one, producing an excess of early counts. The variance correction is
\begin{equation}
    \mathrm{Var}_{\mathrm{early}}^{(e)}(N) =\frac{2\gamma_m \gamma_w(16\gamma_m^2 + 10 \gamma_m \gamma_w-\gamma_w^2)}{(4\gamma_m+\gamma_w)^4} , \!\!\! 
    \label{eq:variance_sigma_x_excited_early}
\end{equation}
and the Mandel $Q_t^{(e)}$ parameter reads as
\begin{equation}
  \begin{split}
  &\!Q_t^{(e)} \\
  &\!= -\frac{\gamma_w \left[4\gamma_m^2(8\gamma_m t +3)
                      +4\gamma_m \gamma_w(2\gamma_m t+3) +\gamma_w^2\right]}
                       {(4\gamma_m+\gamma_w)^2                       \left[\gamma_w+2\gamma_m \left(1+(4\gamma_m+\gamma_w) t \right) \right]}.
                       \!\!\!
    \end{split}
  \label{eq:Q_sigma_x_excited_early}
\end{equation}
At weak dephasing, $\gamma_m\to0$, the pump vanishes and the excited qubit emits a single photon before relaxing to the ground state, so $Q_t^{(e)}\to-1$, the same near-deterministic limit reached when the dephasing is aligned with the energy basis. Here, this limit is approached instead by switching off the dephasing noise. 
The transient contributions in Eqs.~\eqref{eq:mean_with_correc} and~\eqref{eq:var_with_correc} are constants: they are evaluated for a fixed initial state $|\rho(0)\rangle\!\rangle$ and represent the offset that the transient relaxation imprints on the long-time counting statistics. For the mean, this contribution vanishes when the system is prepared in the steady state, $|\rho(0)\rangle\!\rangle = |\rho_{\rm ss}\rangle\!\rangle$ (see Eq.~\eqref{eq:2level_density_ss}), 
since there is then no transient to relax through. In this case the mean correction exactly vanishes, $\langle N\rangle_{\rm early}^{(ss)}=0$, and the leading-order LDT prediction reproduces the mean number of jumps exactly. The variance correction, however, does not vanish, but converges to the small residual offset,
\begin{equation}
    \mathrm{Var}_{\mathrm{early}}^{(ss)}(N) = \frac{8\gamma_m^2 \gamma_w^2}{(4\gamma_m+\gamma_w)^4} ,
    \label{eq:variance_sigma_x_ss}
\end{equation}
giving a Mandel parameter $Q_t^{(ss)}$ with
\begin{equation}
   Q_t^{(ss)} = -\frac{4 \gamma_m \gamma_w}{(4 \gamma_m + \gamma_w)^3 \, t}
   \left[ (4\gamma_m+\gamma_w)t-1 \right] .
\end{equation}
This residual is the stationary boundary contribution to the variance discussed in Sec.~\ref{sec:TheoreticalMethods}.

The excellent agreement between the analytical expressions and the simulations demonstrates that these asymptotic expansions capture the relevant dynamics over the parameter regimes considered. The counting window used in the simulations is sufficiently long to reach the regime where the expansions apply, $t \gg \tau_{\rm relax}$, with $\tau_{\rm relax}$ set by the slowest mode of $\hat{\mathcal{L}}_0$ populated by the initial state. At $\theta=0$, both $\hat{\sigma}_x$ and the Hamiltonian preserve the populations, so the diagonal preparations used here remain within the population sector and relax at the single rate $\Gamma = 4\gamma_m + \gamma_w$. Outside their range of validity, the truncated expansions can exhibit pathologies, such as a negative variance. For the ground state, $\mathrm{Var}_{\mathrm{LDT}}(N)_t + \mathrm{Var}^{(g)}_{\rm early}(N)$ becomes negative for $t < t^\ast = -\mathrm{Var}^{(g)}_{\rm early}(N)/v_{\rm ss}$, but
\begin{equation}
  \Gamma\, t^\ast
  = \frac{16\gamma_m^2 - 2\gamma_m\gamma_w + \gamma_w^2}
         {16\gamma_m^2 + 4\gamma_m\gamma_w + \gamma_w^2} \;<\; 1 ,
  \label{eq:tstar_2level}
\end{equation}
so the apparent negativity occurs only at times shorter than the counting window relevant for the comparison with the simulations. 

\begin{figure}[htbp]
  \centering
  \includegraphics[width=0.45\textwidth]{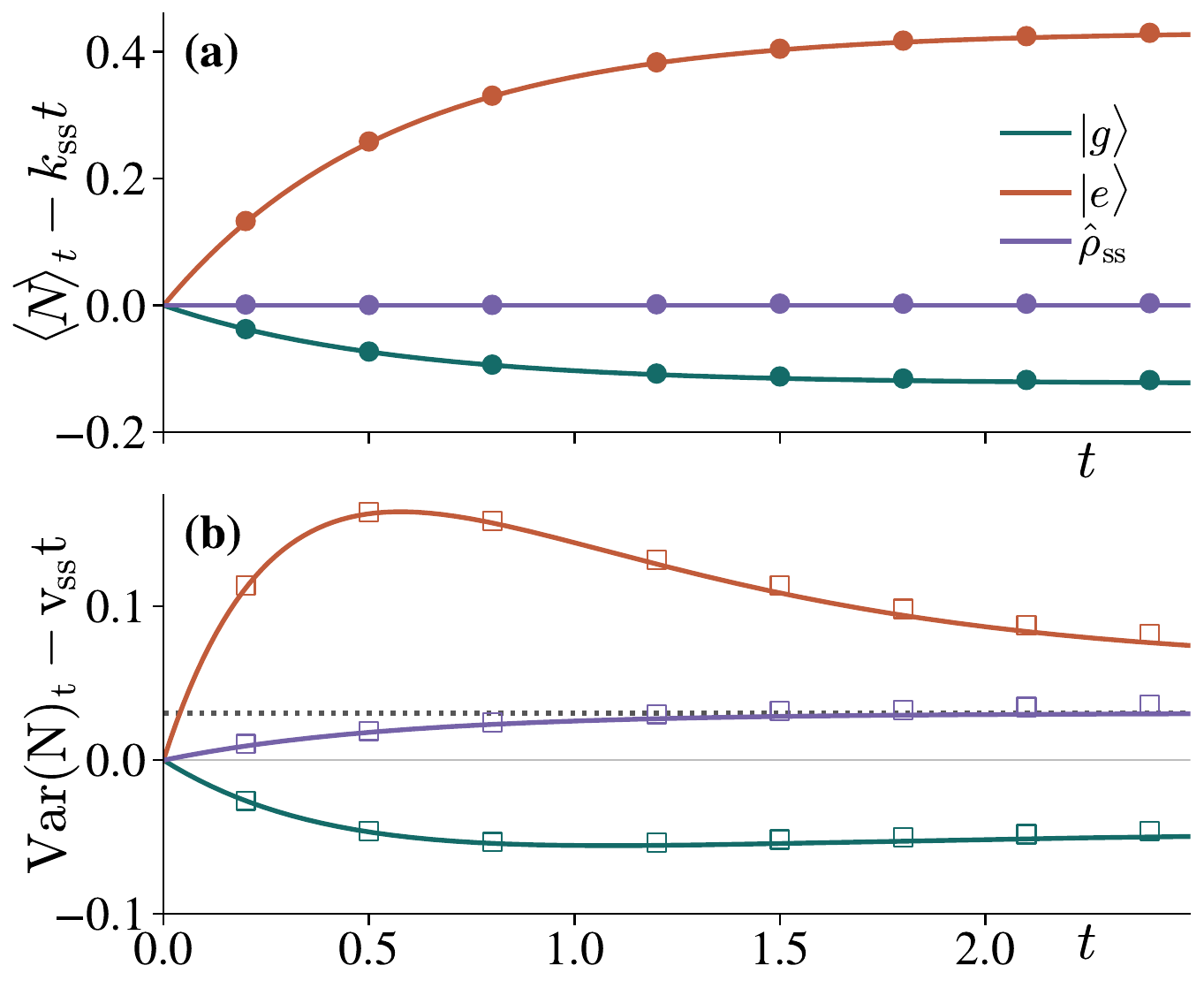} 
  \caption{Deviation of (a) the mean and (b) the variance of the   photon count from their leading-order LDT growth, as functions of the observation time for the two-level emitter with   $\hat X = \hat\sigma_x$. Solid lines: closed-form finite-time   expressions of Eqs.~\eqref{eq:app_ft_mean2lvl} and~\eqref{eq:app_ft_var2lvl}; symbols: quantum-trajectory simulations   ($10^5$ trajectories). Shown are preparations in the ground state $|g\rangle$, the excited state $|e\rangle$, and the steady state $\hat{\rho}_{\rm ss}$ (see legend). After a transient, each curve saturates to its preparation-dependent constant: Eqs.~\eqref{eq:mean_sigma_x_ground_early} and~\eqref{eq:variance_sigma_x_ground_early} for the ground state and Eqs.~\eqref{eq:mean_sigma_x_excited_early} and~\eqref{eq:variance_sigma_x_excited_early} for the excited state. The steady-state preparation is unbiased in the mean, while its variance offset saturates to the boundary term of Eq.~\eqref{eq:variance_sigma_x_ss} (dotted line). Parameters: $\gamma_m = 0.2$, $\gamma_w = 1$.}
  \label{fig:2level_evolution_mean_variance}
\end{figure}

Beyond the asymptotic regime, the mean and variance can be obtained in closed form at \textit{all} times (see Appendix~\ref{app:time_resolved}). Subtracting the leading-order growth, $\mathcal{O}_{\rm mean}(t) = \langle N \rangle_t - k_{\rm ss}\,t$ and $\mathcal{O}_{\rm var}(t) = \mathrm{Var}(N)_t - v_{\rm ss} \,t$, isolates the subextensive part throughout the transient and shows how the initial-state contributions build up in time. Figure~\ref{fig:2level_evolution_mean_variance} shows this for the two-level emitter. At early times the cumulants deviate from their asymptotic growth. As the system relaxes, these deviations saturate to preparation-dependent constants that constitute the initial-state contributions central to this work.

\subsection{Three-level system}

We extend now our formalism to a three-level system. The Hamiltonian of the system is given by
\begin{equation}
    \hat{H} =  \Omega_m \ket{m}\bra{m} +  \Omega_e \ket{e}\bra{e} \,,
\end{equation}
where $\ket{e}$ and $\ket{m}$ are the excited states of the three-level quantum system, with $\Omega_m > \Omega_e$. We assume that $\Omega_m - \Omega_e$ is experimentally accessible, whereas $\Omega_e$ is not directly accessible, corresponding to losses to a reservoir via unmonitored fluorescence emission at rate $\gamma_e$. The fluorescence emission from the level $\ket{m}$ to the level $\ket{e}$ at rate $\gamma_w$ is measured as the clock transition through photodetection (see Fig.~\ref{fig:setup}). Together with the dephasing operator $\hat{X}$, the tilted Liouvillian of our three-level system reads as follows:
\begin{equation}
   \begin{split}
    \hat{\mathcal{L}}_s[\hat{\rho}] &= -i [\hat{H}, \hat{\rho}] -\gamma_m [\hat{X}, [\hat{X},\hat{\rho}]] \\
    &+ \gamma_e \left[ \hat{L}_e \hat{\rho}\hat{L}_e^\dagger - \frac{1}{2} \left( \hat{\rho} \hat{L}_e^\dagger \hat{L}_e +  \hat{L}_e^\dagger \hat{L}_e  \hat{\rho}  \right) \right] \\
    &+ \gamma_w \left[ e^{-s} \hat{L}_w  \hat{\rho}\hat{L}_w ^\dagger - \frac{1}{2} \left( \hat{\rho} \hat{L}_w ^\dagger \hat{L}_w  +  \hat{L}_w ^\dagger \hat{L}_w   \hat{\rho}  \right) \right] ,
    \end{split}
\label{eq:lindblad_threelevel}
\end{equation}
where $\hat{L}_w = \ket{e}\bra{m}$ and $\hat{L}_e = \ket{g}\bra{e}$. It remains to specify the operator $\hat{X}$ through which the dephasing noise acts. As in the two-level case, where $\theta$ set the orientation of $\hat X$, we introduce a one-parameter family. Let
\begin{align}
    |\phi_1(\kappa)\rangle &= \frac{1}{\sqrt{1+\kappa^2}} \left( |e\rangle + \kappa |g\rangle \right), \\
    |\phi_2(\kappa)\rangle &= \frac{1}{\sqrt{2}} \left( \sqrt{1-\kappa^2}\,|m\rangle - \kappa |e\rangle + |g\rangle \right),
\end{align}
with $|\phi_3(\kappa)\rangle$ completing the orthonormal basis, and define
\begin{equation}
    \hat{X}(\kappa) = |\phi_1(\kappa)\rangle\langle \phi_1(\kappa)| - |\phi_2(\kappa)\rangle\langle \phi_2(\kappa)| ,
\end{equation}
which annihilates $|\phi_3\rangle$.

\begin{figure}[htbp]
   \includegraphics[width=0.45\textwidth]{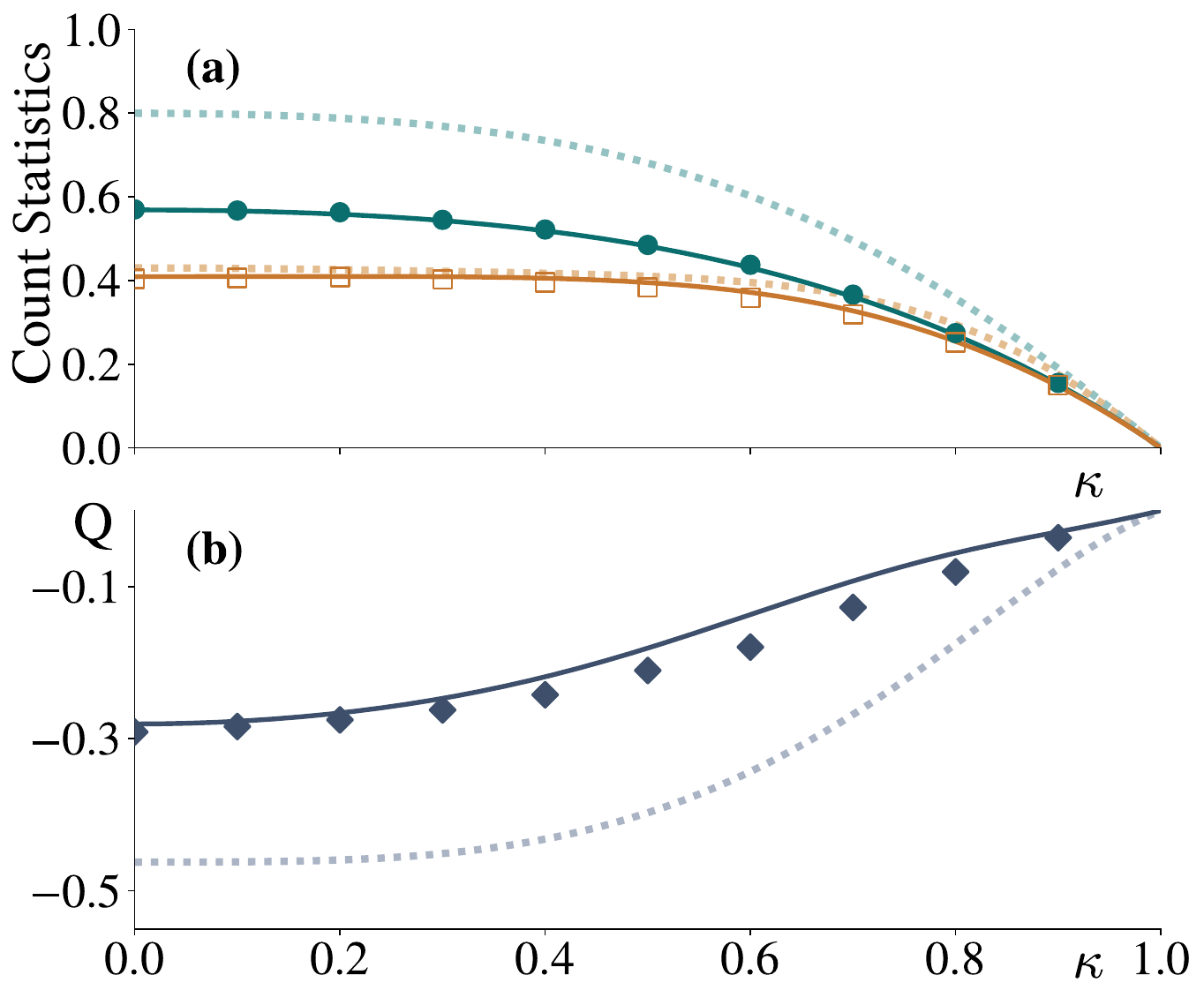}
   \caption{
    (a) Mean number (green) and variance (orange) of clicks as functions of $\kappa$ for the observable $\hat{X}(\kappa)$. 
    (b) Mandel $Q_t$ parameter (dark blue) as a function of $\kappa$. 
    In both panels, dashed lines denote the LDT predictions, while solid lines include both the LDT and transient contributions arising from the initial state (Eqs.~\eqref{eq:app_mean_summary} and~\eqref{eq:app_var_summary} evaluated analytically). Scatter points correspond to numerical simulations performed using the Kraus-operator formalism (see Appendix~\ref{app:Kraus_Formalism}). Parameters: $\Omega_m = 1$, $\Omega_e = 0.1$, $\gamma_w = 3$, $\gamma_e = 4$, $\gamma_m = 6$. Simulations were performed with time step $dt = 0.001$ over $1000$ steps using $10^5$ trajectories. Initial state $\hat{\rho}(0) = \ket{e}\bra{e}$. }
   \label{fig:3level_mean_variance_Q_excited}
\end{figure}

The parameter $\kappa$ controls how strongly the noise couples to the emission cycle. After each emission the system is left in $|e\rangle$ and cannot emit again until it has returned to $|m\rangle$ through $|e\rangle\to|g\rangle\to|m\rangle$, the last step being driven by the $\hat X$-dephasing channel. Waiting for several comparable steps in sequence is more predictable than waiting for a single random one: at $\kappa=0$, where $|\phi_1\rangle=|e\rangle$, the noise acts directly on the post-emission state and drives the return to $|m\rangle$ efficiently, so no single step dominates the waiting time between clicks; the emission is then more regular than a memoryless process and the statistics are sub-Poissonian. As $\kappa\to1$ the noise connects $|g\rangle$ back to $|m\rangle$ ever more weakly: the re-pumping becomes the slowest step and increasingly dominates the waiting time between clicks, causing the emission rate to collapse and the statistics to approach the Poissonian limit (see Fig.~\ref{fig:3level_mean_variance_Q_excited}). At $\kappa=1$ the noise no longer acts on $|m\rangle$ at all and the emission ceases. The parameter $\kappa$ thus plays for the cascade the role that $\theta$ played for the two-level emitter.

The uncounted decay also sets when the asymptotic expressions become accurate. In the two-level emitter at $\theta=0$, for instance, the relevant relaxation was fast: a diagonal preparation relaxed at $\Gamma = 4\gamma_m + \gamma_w$. In the cascade, relaxation must instead proceed through all three steps of the cycle, including the uncounted transition. The relaxation time is therefore controlled by a process invisible to the detector and can become longer when $\gamma_e$ is small. An observation window comfortably long in the two-level case may here be too short.

\section{Figures of merit from temporal statistics}
\label{sec:figuresofmerit}

\subsection{Precise timekeeping in the transient regime} 
\label{sec:timekeeping}

If the detected photons are regarded as the ticks of a clock, the counting statistics translate directly into the standard figures of merit of timekeeping in autonomous quantum clocks~\cite{Meier2023,erker_autonomous_2017,Manikandan2023}.
To begin, we can map our protocol, in which we fix an observation window $t$ and count the number of ticks $N$ within it, to a waiting-time description, in which one records the times $\tau_1, \tau_2, \tau_3, \dots$ between successive ticks~\cite{Meier2023}. Two conditions are worth stating for this identification to be exact. First, every detected photon resets the system to the same fixed state (to $\ket{g}$ for the qubit), guaranteeing that all inter-tick waiting times after the first one are i.i.d., with only the first waiting time depending on the initial state. Second, the clock ticks with certainty.

With that in mind, two of these figures of merit are fixed by the leading-order rates alone~\cite{erker_autonomous_2017}: the \emph{resolution}, i.e., how finely the clock divides time, is given by the stationary tick rate $\nu = k_{\rm ss}$, while the \emph{accuracy}, i.e., the number of ticks produced before the accumulated timing error reaches one tick period, is $N_{\rm acc} = k_{\rm ss}/v_{\rm ss} = 1/(1+Q_{\rm ss})$, with $Q_{\rm ss} = v_{\rm ss}/k_{\rm ss}-1$ built from the time-extensive rates alone. A sub-Poissonian tick stream, $Q_{\rm ss}<0$, is thus a clock more accurate than one limited by shot noise. The two metrics obey the trade-off $N_{\rm acc}\,\nu^2 \le \gamma_w^2$~\cite{Meier2023}, which in our variables reads $k_{\rm ss}^3 \le \gamma_w^2 v_{\rm ss}$. Note that the initial-state transient terms neither improve nor evade the accuracy-resolution trade-off, which constrains the rates only. This distinction clarifies the point $\theta=\frac{\pi}{2}$ in Fig.~\ref{fig:2level_mean_variance_Q_excited}. There both stationary rates, $k_{\rm ss}$ and $v_{\rm ss}$, vanish, so no stationary tick process exists and the resolution and accuracy defined above are undefined. Approaching this point, the rates stay finite, yet $Q_t$ turns strongly negative through the transient constants. This is to be understood as the near-deterministic first emission given sufficient time, not a gain in clock accuracy.

The initial-state transient terms instead show up in the figures of merit that refer to a finite record. From the expansions
\begin{equation}
    \langle N \rangle_t = k_{\rm ss} t + c_1, \quad \mathrm{Var}(N)_t = v_{\rm ss} t + c_2,
\end{equation}
where $c_1 = -(\ln c_{\rm dom})'(0)$ and $c_2 = (\ln c_{\rm dom})''(0)$, consider the estimator $\hat t = N/k_{\rm ss}$, which infers the elapsed time directly from the observed count. This is a deliberately simple choice rather than an optimal one~\cite{Prech2025}, as we are interested in the constant \emph{bias} it acquires,
\begin{equation}
  t_{\rm bias} = \langle \hat t\rangle - t = \frac{c_1}{k_{\rm ss}} ,
\end{equation}
independent of the observation time: the tick train is shifted by a fixed offset, so the clock reads systematically ahead or behind depending on the preparation. In the two-level model, a preparation in $|e\rangle$ can emit immediately, giving the tick train a head start ($c_1>0$), whereas a preparation in $|g\rangle$ must first be excited, delaying the train ($c_1<0$). The steady state $\hat{\rho}_{\rm ss}$ gives exactly $c_1=0$, i.e., it is unbiased. Unlike the metrics below, this offset persists in the asymptotic limit.

The Fano factor $F(t) = \mathrm{Var}(N)_t/\langle N\rangle_t$ approaches its stationary value $F_\infty = v_{\rm ss}/k_{\rm ss} = 1/N_{\rm acc}$ with a preparation-dependent transient
\begin{equation}
  F(t) = F_\infty + \frac{c_2 - F_\infty c_1}{k_{\rm ss}t} + \mathcal{O}(t^{-2}) ,
\end{equation}
so that any accuracy inferred from a finite record is shifted from $N_{\rm acc}$ by the transient constants, an offset that decays as $t\to \infty$.

Finally, the relative root-mean-square error of the estimator $\hat t$,
\begin{equation}
  \frac{\sqrt{\langle (\hat{t}-t)^2 \rangle}}{t}
  = \frac{\sqrt{v_{\rm ss}\,t + c_2 + c_1^2}}{k_{\rm ss}\,t }
  \;\xrightarrow[t\to\infty]{}\;
  \sqrt{\frac{F_\infty}{k_{\rm ss}\,t}} ,
\end{equation}
approaches the standard $1/\sqrt{t}$ behavior in the long-time limit. Both transient constants contribute additively to the numerator, with $c_1^{2}$ representing the squared bias. The Fano factor and relative error therefore approach their asymptotic behavior along a preparation-dependent transient set by $c_1$ and $c_2$, with only the bias persisting.

\subsection{Implications for quantum sensing}
Given the progressively increasing interest in using quantum clocks as sensors in various avenues~\cite{Qsensing,kawasaki_quantum_2025,nichol_elementary_2022,Klaus1,Klaus2}, including tests of fundamental physics~\cite{Prlessay,zych_quantum_2011,verma_effect_2021,qhj9-pc2b,Pikovski2,bothwell_resolving_2022}, it is interesting to consider the quantum sensing prospects of autonomous quantum clocks as described. To this end, we consider an exemplary scenario where the autonomous quantum emitter functions as a quantum sensor for an external coherent field, of amplitude $\gamma_d$ and frequency $\omega_d$, driving the two-level emitter of transition frequency $\Omega$ on top of the dephasing noise fueling the clock. Working in the frame rotating at $\omega_d$ and within the rotating-wave approximation (see Appendix~\ref{app:driven} for details), the counting statistics of the driven emitter, including the fluorescence measurement, follow from the tilted Lindbladian,
\begin{equation}
    \begin{split}
    \hat{\mathcal{L}}_s[\hat{\rho}] &= -i [\hat{H}_d^{\mathrm{RWA}}, \hat{\rho}] -\gamma_m [\hat{X}, [\hat{X},\hat{\rho}]] \\
    &+ \gamma_w \Big[  e^{-s} \, \hat{\sigma}_- \hat{\rho} \, \hat{\sigma}_+ - \frac{1}{2} \left( \hat{\rho} \, \hat{\sigma}_+  \hat{\sigma}_- +  \hat{\sigma}_+ \hat{\sigma}_- \, \hat{\rho}  \right) \Big] ,
   \end{split}
\end{equation}
where
\begin{equation}
    \hat{H}_d^{\mathrm{RWA}} = \Lambda(\theta) \ket{e}\bra{e} + g(\theta) \hat{\sigma}_x ,
\end{equation}
with an effective splitting $\Lambda(\theta) = \Omega- \omega_d + 2\gamma_d\sin\theta$ and coupling $g(\theta) = \gamma_d\cos\theta$. The operator $\hat{X}(\theta)$ is the one introduced in Eq.~\eqref{eq:X_2level}, and is understood in the rotating frame.

The relevant figure of merit for quantum sensing is the Fisher information~\cite{Fisher1922} of the counting statistics of clock ticks, with the amplitude of the external drive $\gamma_d$ as the parameter to be estimated.  As shown in Appendix~\ref{app:driven}, the transient, initial-state contributions that markedly improve the description of $\langle N \rangle_t$ and $\mathrm{Var}(N)_t$ do enter the Fisher information, but only at the same order as every other subleading correction. This improvement hence does not translate into a distinguishable gain in estimation precision.

The sensing precision is therefore largely determined by steady-state quantities alone, through $\partial_{\gamma_d}k_{\rm ss}$ and $v_{\rm ss}$. While dephasing degrades the estimation of $\gamma_d$, at resonance and $\theta=0$, photon counting can still be near-optimal within the regime $\gamma_m\gamma_w/2 \ll \gamma_d^{2} \ll \gamma_w^{2}/8$, a window that closes as dephasing grows (Appendix~\ref{app:driven}). In the absence of dephasing and at long times, photon counting saturates the \textit{quantum} Fisher information in the weak-drive limit, where the emitted photons carry all the available information about $\gamma_d$.

\section{Conclusions}
We considered simple models to describe the temporal emission statistics of an autonomous quantum coherent emitter, viewed as a quantum clock in the transient regime, as the emitter evolves towards a steady state with quantum coherences. The transient contributions we evaluated go beyond the leading-order large deviation theory and describe how the temporal statistics of the quantum emitter are influenced by its initial conditions as well as by coherences in its steady state. Our results were obtained using two complementary methods: (i) stochastic trajectory simulations within the Kraus-operator formalism, and (ii) closed-form analytical results derived from a perturbative expansion of the tilted Lindbladian generator. Our findings were also numerically benchmarked against the exact eigenvectors of the full tilted Lindbladian.

Applying large deviation theory, and comparable alternatives~\cite{Jordan1,sukhorukov_conditional_2007}, to the finite-time counting statistics of quantum emitters is fruitful in that it gives exact results for a difficult counting problem that can potentially involve multiple reservoirs. This detailed open-quantum description is, to some extent, lacking in the atomic-clock literature, where the focus is often on characterizing frequency/phase stability through measurements in the semiclassical ($\langle N\rangle_t \gg 1$)  regime of probing the laser. Nevertheless, the short-time dynamics and the few-quanta ($\langle N\rangle_t\sim 1$) regime offer a complementary perspective on the figures of merit for a timekeeping device, with important ramifications for understanding the fundamental limitations to precise timekeeping in the quantum regime.

We primarily discussed the implications for precise timekeeping in the quantum regime and described the relevant figures of merit.
We also discussed the implications for quantum sensing using artificial atomic emitters. When a quantum clock is used as a sensor, precise knowledge of its figures of merit is essential, especially at the short timescales of most experiments. Motivated by this, we presented the example of an autonomous quantum clock subject to a driving field, and discussed the optimal regime of operation for it to function as a quantum sensor. 

While there are important links to the physics of atomic clocks, as noted, we primarily focused on the regime of autonomous quantum clocks described by simple models fueled by elementary quantum thermodynamic resources. Extending our considerations to real atomic clocks by including the spectral diffusion of an ensemble of autonomous quantum coherent emitters and accounting for their feedback stabilization using real atoms invites considerable further research. We defer a detailed analysis of these questions to future work. 

\section{acknowledgements}  

OA is partially supported by the Knut and Alice Wallenberg Foundation (2023.0256) through the Wallenberg Centre for Quantum Technology (WACQT). SKM acknowledges support from the Department of Atomic Energy, Government of India, under Project Identification No. RTI4007. OA made leading contributions to deriving the transient counting statistics, and SKM to conceptualizing the research.  Both authors contributed equally to analyzing the results and writing the manuscript. The authors acknowledge Claude (Fable 5) for suggesting the perturbative approach from Refs.~\cite{FlindtNovotnyBraggio2010}, used in Appendix~\ref{app:pert_expansion}, which clarified for us the physical origin of the transient contributions.

\noindent\textit{Data availability statement.---}  A Mathematica notebook containing the analytic calculations using the tilted Lindbladian presented in the manuscript can be found at the following \href{https://github.com/OscarArandesTejerina/two_level_counting_statistics.git}{GitHub repository link}.

\appendix
\section{Kraus Operators Formalism} \label{app:Kraus_Formalism}
To benchmark our theoretical predictions, we perform numerical simulations using the Kraus-operator formalism~\cite{kraus1983}, which has been extensively employed in studies of fluorescence emission~\cite{Jordan2016,Manikandan2019,Lewalle2020,Manikandan2023,Benny_Manikandan}. The analysis relevant at the large time limit without coherences has been the focus of earlier works~\cite{Manikandan2023,Benny_Manikandan} (for exemplary codes, see \footnote{\label{footSim}See the following~ \href{https://github.com/sreenathkm92/clocks.git}{ Github repository link 1} for simulations in Ref.~\cite{Manikandan2023} and \href{https://github.com/sreenathkm92/QuantumEmitter}{ Github repository link 2} for simulations in Ref.~\cite{Benny_Manikandan}} for links to the corresponding numerical simulations). Our theoretical and numerical analysis in the present work extends to the transient regime, considering generalized dephasing noise that could lead to coherences in the steady state as well. Importantly, this numerical approach is independent of the LDT framework used in the main text and provides an alternative method for validating our theoretical results.

First, consider a three-level system subject to fluorescence measurement associated with the decay transition of the levels $|m\rangle$ to $|e\rangle$. The conditional evolution of the system during an infinitesimal time interval $dt$ can be modeled within the quantum trajectory formalism using Kraus operators~\cite{Wiseman_Milburn_2009,Nielsen_Chuang_2010}. These operators account for the measurement backaction associated with the two possible outcomes of the fluorescence measurement: either a photon is detected (click) or no photon is detected (no-click). Conditioned on the measurement outcome $j\in\{0,1\}$, the density operator evolves according to
\begin{equation}
    \hat{\rho} (t + dt) = \frac{\hat{M}_w(j) \, \hat{\rho}(t) \, \hat{M}_w(j)^\dagger}{ \mathrm{tr} \left( \hat{M}_w(j) \, \hat{\rho}(t) \, \hat{M}_w(j)^\dagger\right)},  
\end{equation}
where the Kraus operators $\hat{M}_w (0)$ and $\hat{M}_w(1)$ represent a no-click and a click in the detector. A formal derivation of the exact expression of these operators is given in Appendix~\ref{app:Kraus_Fluorescence} (see also Refs.~\onlinecite{Lewalle2020, Manikandan2023} for a phenomenological derivation). They are given by
\begin{equation}
   \begin{split}
     \hat{M}_w(0) &=
     \begin{pmatrix}
     \sqrt{1-\epsilon_w} & 0 & 0 \\
     0 & 1 & 0 \\
     0 & 0 & 1
    \end{pmatrix},
    \\[1ex]
    \hat{M}_w(1) &=
    \begin{pmatrix}
    0 & 0 & 0 \\
    \sqrt{\epsilon_w} & 0 & 0 \\
    0 & 0 & 0
    \end{pmatrix} ,
  \end{split}
\end{equation}
in the ordered basis $\{\ket{m},\ket{e},\ket{g}\}$, where $\epsilon_w = \gamma_w \, dt$. The probability of obtaining a click is given by Born's rule, according to the standard axioms of quantum mechanics,
\(
p_{\mathrm{click}} = \operatorname{tr} \left( \hat{M}_w(1) \, \hat{\rho}(t) \, \hat{M}_w(1)^\dagger \right),
\)
while the probability of no click is
\(
p_{\mathrm{no-click}} = \operatorname{tr} \left( \hat{M}_w(0) \, \hat{\rho}(t) \, \hat{M}_w(0)^\dagger \right),
\)
with
\(
p_{\mathrm{click}} + p_{\mathrm{no-click}} = 1. 
\)
It is straightforward to verify that $\hat{M}_w(0)^{\dagger} \hat{M}_w(0)+\hat{M}_w(1)^{\dagger} \hat{M}_w(1)=\hat{\mathds{1}}_{3\times 3}$, such that these measurement operators form a positive operator-valued measure (POVM). 

Similarly, one can model the inaccessible losses in the system due to spontaneous emission between levels $\ket{e}$ to $\ket{g}$ using the Kraus operators in the ordered basis \(\{\ket{m}, \ket{e}, \ket{g}\}\):
\begin{equation}
   \begin{split}
    \hat{M}_e(0) &= 
    \begin{pmatrix}
     1 & 0 & 0 \\
     0 & \sqrt{1-\epsilon_e} & 0 \\
     0 & 0 & 1
    \end{pmatrix}, \\[1ex]
     \hat{M}_e(1) &= 
    \begin{pmatrix}
     0 & 0 & 0 \\
     0 & 0 & 0 \\
     0 & \sqrt{\epsilon_e} & 0
    \end{pmatrix}, 
    \end{split}
\end{equation}
where $\epsilon_e = \gamma_e dt$ and $\gamma_e$ is the rate of spontaneous emission into the environment. Note that $\ket{e} \to \ket{g}$ emissions correspond to \textit{unobserved} events. The quantum evolution is therefore obtained by averaging the state update over both measurement outcomes:
\begin{equation}
    \hat{\rho} (t + dt) = \frac{\sum_{j=0,1} \hat{M}_e(j) \, \hat{\rho}(t) \, \hat{M}_e(j)^\dagger}{ \mathrm{tr} \left( \sum_{j=0,1}\hat{M}_e(j) \, \hat{\rho}(t) \, \hat{M}_e(j)^\dagger\right)} .
\end{equation}
In addition to the fluorescence measurement, the dephasing noise introduced in the main text can be realized as a continuous weak measurement of $\hat{X}$ whose outcome is left unread. Averaging over that outcome reproduces the dephasing dynamics (double-commutator term below). For concreteness, and because it is the form used in our simulations, we describe this realization explicitly via its Kraus operators. In this case, the corresponding Kraus measurement operators are given by (see Appendix~\ref{app:Kraus_Continuous} for a separate derivation):
\begin{equation}
    \hat{M}_{\hat{X}}(r) =  \left( \frac{4 \epsilon_m}{\pi} \right)^{\frac{1}{4}}  e^{-2\epsilon_m (r \hat{\mathds{1}}-\hat{X})^2},
\end{equation}
where $\epsilon_m = \gamma_m \, dt$, and the measurement operators satisfy the completeness relation $\int dr \hat{M}_{\hat{X}}^\dagger(r) \hat{M}_{\hat{X}}(r) = \hat{\mathds{1}}$. 

Finally, we can write the full dynamics associated with the discrete fluorescence measurement outcomes $\hat{M}_w(k)$ and the unconditional continuous monitoring over an infinitesimal interval $dt$, described by $\hat{M}_{\hat{X}}(r)$, while also accounting for inaccessible losses through $\hat{M}_e$. Defining 
\begin{equation}
   \tilde{\rho}(t+dt) = \int dr\, \hat{\mathcal M}_{dt}\, \hat{\rho}(t)\, \hat{\mathcal M}_{dt}^\dagger ,
\end{equation}
where $ \hat{\mathcal{M}}_{dt} \equiv \hat{M}_{\hat{X}}(r) e^{-i\hat{H}dt}$ also accounts for the Hamiltonian evolution, the conditional state update is 
\begin{equation}
\begin{split}
    &\hat{\rho}(t+dt) \\
    &= \sum_{j=0,1} \hat{M}_e(j)\, 
    \frac{\hat{M}_w(k)\,\tilde{\rho}(t+dt)\,\hat{M}_w^\dagger(k)
    }{
    \mathrm{tr}\!\left[ \hat{M}_w(k)\, \tilde{\rho}(t+dt)\, \hat{M}_w^\dagger(k) \right]}
     \hat{M}_e^\dagger(j) ,
\end{split}
\end{equation}
conditioned on observing a click $(k=1)$ or no click $(k=0)$. For the simulations, we approximate
\begin{equation}
   \begin{split}
    &\int dr \hat{\mathcal{M}}_{dt} \, \hat{\rho}(t) \, \hat{\mathcal{M}}_{dt}^\dagger \\
    &\approx \hat{\rho}(t)  
    -i [\hat{H}, \hat{\rho}(t)] \, dt 
    - \gamma_m [\hat{X},[\hat{X},\hat{\rho}(t)]] \, dt.
   \end{split}
\end{equation}
Lastly, a click is recorded with probability
\begin{equation}
  p_{\mathrm{click}} = \mathrm{tr} \left[
  \hat{M}_w(1) \,  \tilde{\rho}(t+dt) \,  \hat{M}_w^\dagger(1)
\right].
\end{equation}

For the two-level system, note that there are no inaccessible losses. Otherwise, the structure for the dynamics is analogous. With this formalism, we can simulate and characterize the fluorescence detector clicks through the average number of clicks observed in a finite duration $t$, $\langle N \rangle_t$, and its variance, $\langle N^2 \rangle_t - \langle N \rangle^2_t$, as presented in the main text.

\section{Kraus Operators for Fluorescence Emission from $|m\rangle$ to $|e\rangle$} \label{app:Kraus_Fluorescence}
Here we derive the corresponding Kraus operators associated with fluorescence emission, specifically for transitions from the state $|m\rangle$ to $|e\rangle$ in the three-level system (the two-level case follows analogously upon relabeling the states).

Consider the Hilbert space $\mathcal{H} = \mathcal{H}_{\mathrm{S}} \otimes \mathcal{H}_{\mathrm{field}}$, where the $m\text{-}e$ subspace of the system $S$ is coupled to an electromagnetic field mode. We assume that the system is prepared in the arbitrary state
$|\psi_{0}\rangle$ while the field mode is initially in the vacuum state $\ket{0}$. The field mode is the one that carries the fluorescence emitted by the system into a waveguide. The interaction Hamiltonian is then given by~\cite{Benny_Manikandan}
\begin{equation}
    \hat{H}_{\mathrm{int}} \, dt =  \sqrt{\gamma_w dt}  \, \left( \hat{\sigma}_+ \otimes \hat{a} + \hat{\sigma}_- \otimes \hat{a}^\dagger \right),
\end{equation}
where the system raising and lowering operators are defined as $\hat{\sigma}_+ = \ket{m}\bra{e}$ and $\hat{\sigma}_- = \ket{e}\bra{m}$, and $\hat{a}$, $\hat{a}^\dagger$ are the annihilation and creation operators of the output field. Note that we have implicitly assumed the rotating wave approximation. The composite state after a time $dt$ is given by
\begin{equation}
  \begin{split}
    &\ket{\psi_{dt}} = e^{-i \hat{H}_{\mathrm{int}} dt} \ket{\psi_0} \otimes \ket{0} \\
    &\approx 
    \bigg[ \hat{\mathds{1}} - i \sqrt{\gamma_w dt}  \, \left( \hat{\sigma}_+ \hat{a} + \hat{\sigma}_- \hat{a}^\dagger \right) \\
    &- \frac{\gamma_w dt}{2} \Big(\! \ket{m}\bra{m}(\hat{a}^\dagger \hat{a} +1)  + \ket{e}\bra{e} \hat{a}^\dagger \hat{a} \Big)  \bigg] 
    \! \ket{\psi_0} \otimes \ket{0}, \!\!\!\!\!\!
  \end{split}
\end{equation}
where we have expanded the time-evolution operator to linear order in $dt$. The measurement operator corresponding to a null event (no detection event) is then given by
\begin{equation}
    \begin{split}
     &\hat{M}_w(0)\ket{\psi_{0}} = \braket{0|\psi_{dt}} \\
     &=  \bigg[ (1 - \frac{\gamma_w dt}{2}) \ket{m}\bra{m} + \ket{e}\bra{e} + \ket{g}\bra{g}  \bigg] \ket{\psi_{0}} \\
     &\approx \bigg[ \sqrt{1 - \gamma_w dt} \ket{m}\bra{m} + \ket{e}\bra{e} + \ket{g}\bra{g} \bigg] \ket{\psi_{0}} .
     \end{split}
\end{equation}
Analogously, for a detection event we have
\begin{equation}
     \hat{M}_w(1)\ket{\psi_{0}} = \braket{1|\psi_{dt}} = \left( -i  \sqrt{\gamma_w dt} \ket{e}\bra{m} \right) \ket{\psi_{0}} ,
\end{equation}
which, up to an irrelevant global phase, corresponds to the Kraus operator stated in Appendix~\ref{app:Kraus_Formalism}.

\section{Kraus Operator for Continuous Quantum Measurements} \label{app:Kraus_Continuous}

This appendix derives the Kraus operators associated with a continuous measurement of $\hat{X}$, which implements the dephasing dynamics considered in the main text. Continuously monitored quantum systems, in which information about the system is acquired through measurement, are naturally described within the framework of \textit{continuous quantum measurement}. For completeness, we briefly review the relevant formalism and fix notation, following Refs.~\onlinecite{Jacobs2006, Manikandan2023}.

Consider a Hilbert space $\mathcal{H} = \mathcal{H}_{\mathrm{S}} \otimes \mathcal{H}_{\mathrm{M}}$. The system of interest is coupled to an auxiliary system, referred to as the \textit{meter}, which is, in turn, subjected to a von Neumann measurement. The meter is a continuous-variable system, $\mathcal{H}_{\mathrm{M}}$, with operators $\hat{x}$ and $\hat{p}$ satisfying $[\hat{x}, \hat{p}] = i$. The Hermitian operator of the system being measured is denoted by $\hat{X}$, with eigenstates $\ket{x_s}$ and corresponding eigenvalues $x_s$, i.e., $\hat{X} \ket{x_s} = x_s \ket{x_s}$.

We prepare the meter in an initial pure Gaussian state
\begin{equation}
    \ket{\psi_M} = \left( \frac{4 \gamma_m}{\pi} \right)^{\frac{1}{4}} \int_{-\infty}^{\infty} dx \, e^{-2\gamma_m x^2} \ket{x}.
\end{equation}
This corresponds to a zero-mean, Gaussian-weighted superposition of eigenstates of the meter operator $\hat{x}$, i.e., $\hat{x}\ket{x} = x\ket{x}$. The constants here are chosen for convenience. Note that the parameter $\gamma_m$ determines the width of the wavefunction, with variance $\propto 1/\gamma_m$. 

Since the meter interacts with the system through an interaction Hamiltonian
\begin{equation}
    \hat{H}_{\mathrm{int}} dt =  \sqrt{dt} \hat{X}\otimes \hat{p},
\end{equation}
the measurement outcome is inferred from shifts in the meter position $\hat{x}$ due to the observable $\hat{X}$ being coupled to the meter momentum $\hat{p}$. The parameter $\gamma_m$ will control the measurement strength: larger $\gamma_m$ corresponds to a more localized state in position, allowing for better distinguishability of measurement outcomes and thus a \textit{stronger} measurement, while smaller $\gamma_m$ leads to a \textit{weaker} measurement.

Defining
\begin{equation}
   \phi(x) = \left(\frac{4\gamma_m}{\pi}\right)^{1/4} e^{-2\gamma_m x^2},
\end{equation}
the state after a time interval $dt$ becomes
\begin{equation}
  \begin{split}
  \ket{\psi_{dt}}
   &=  e^{-i\hat{H}_{\mathrm{int}}dt} \ket{\psi_0}\otimes\ket{\psi_M} \\
   &= e^{-i\sqrt{dt}\,\hat{X}\otimes\hat{p}} \sum_s c_s \ket{x_s}\otimes\ket{\psi_M} \\
   &= \sum_s c_s \ket{x_s} \otimes \int_{-\infty}^{\infty} dx\, \phi(x)\, e^{-i\sqrt{dt}\,x_s\hat{p}} \ket{x} \\
   &= \sum_s c_s \ket{x_s} \otimes \int_{-\infty}^{\infty} dx\, \phi(x)\, \ket{x+\sqrt{dt}\,x_s} ,
\end{split}
\end{equation}
where we have explicitly expanded the initial state of our system in terms of the eigenvectors $\ket{x_s}$ of $\hat{X}$. The Kraus operator corresponding to a given readout $x = q$ is defined by projecting the meter onto the state $\ket{q}$, which corresponds to measuring the position quadrature $\hat{x}$ (homodyne measurement on the meter):
\begin{equation}
   \begin{split}
    &\hat{M}_{\hat{X}}(q) \ket{\psi_0} = (\hat{\mathds{1}}\otimes \braket{q|)|\psi_{dt}} \\
    &=\sum_s c_s \ket{x_s}  \int_{-\infty}^{\infty} dx \, \phi(x) \braket{q|x+\sqrt{dt} \, x_s}  \\
    &= \sum_s c_s \ket{x_s} \left( \frac{4 \gamma_m}{\pi} \right)^{\frac{1}{4}} e^{-2\gamma_m (q-\sqrt{dt}\,x_s)^2} \\
    &= \left( \frac{4 \gamma_m}{\pi} \right)^{\frac{1}{4}}  e^{-2\gamma_m (q \hat{\mathds{1}}-\sqrt{dt}\,\hat{X})^2} \ket{\psi_0} .
     \end{split}
\end{equation}
Rescaling the readout as $r=\frac{q}{\sqrt{dt}}$ we can express the measurement operator as
\begin{equation}
   \begin{split}
    \hat{M}_{\hat{X}}(r) &= \left( \frac{4 \gamma_m dt}{\pi} \right)^{\frac{1}{4}}  e^{-2\gamma_m dt (r \hat{\mathds{1}}-\hat{X})^2} \\
    &= \left( \frac{4 \epsilon_m}{\pi} \right)^{\frac{1}{4}}  e^{-2\epsilon_m (r \hat{\mathds{1}}-\hat{X})^2},
    \end{split}
\end{equation}
where $\epsilon_m = \gamma_m \, dt$.

\section{Vectorization and Liouville-Space Representation} \label{app:vectorization}

In this appendix, we summarize the vectorization procedure used to represent (super) operators in Liouville space~\cite{Gyamfi2020,LandiReview2024}. A density operator $\hat\rho$ acting on a $d$-dimensional Hilbert space $\mathcal{H}$ can be represented as a vector $|\rho\rangle\!\rangle$ in the $d^2$-dimensional Liouville space $\mathcal{H}\otimes\bar{\mathcal{H}}$, where $\bar{\mathcal{H}}$ denotes the complex-conjugate Hilbert space, through the vectorization map
\begin{equation}
  \hat\rho(t) = |\psi\rangle\langle\psi| \;\longrightarrow\; |\rho(t)\rangle\!\rangle = |\psi\rangle\otimes|\psi\rangle^{*} ,
\end{equation}
where $|\psi\rangle^{*}$ denotes the vector obtained by complex-conjugating the components of $|\psi\rangle$ in a chosen orthonormal basis. This specific vectorization map, or, equivalently, the ordering of matrix elements within $|\rho\rangle\!\rangle$, is a matter of convention. Throughout, we adopt the row-stacking convention, in which the rows of the matrix representation of $\hat\rho$ are concatenated into a single column vector $|\rho\rangle\!\rangle$. In this convention, the triple-product identity for three matrices $A,B, C$ reads
\begin{equation}
    |ABC\rangle \! \rangle = (A \otimes C^T)\,|B\rangle \! \rangle,
\end{equation}
where the first factor represents left multiplication, while the second factor represents right multiplication. Using this identity, Hamiltonian terms of the Lindblad equation are vectorized as
\begin{align}
    | H \rho \rangle \! \rangle &= (H \otimes \mathds{1})\,|\rho \rangle \! \rangle, \\
    | \rho H \rangle \! \rangle &= (\mathds{1} \otimes  H^T )\,|\rho \rangle \! \rangle,
\end{align}
while jump terms take the form
\begin{equation}
    | L \rho L^\dagger \rangle \! \rangle = \big(L \otimes (L^\dagger)^T \big)\,|\rho\rangle \! \rangle.
\end{equation}
Dissipative terms such as $L^\dagger L \rho$ and $\rho L^\dagger L$ follow analogously. Note that the placement of the transpose and the order of Kronecker factors is specific to the row-stacking convention; switching to column-stacking would reverse them.

We can then express the Liouvillian of the two-level system, obtained from Eq.~\eqref{eq:lindblad_twolevel}, in its vectorized form as
\begin{equation}
   \begin{split}
    \mathcal{L}_s =& -i \left( H \otimes \mathds{1} - \mathds{1} \otimes H^T  \right) \\
    & - \gamma_m \left(X^2 \otimes \mathds{1} - 2 X\otimes X^T + \mathds{1} \otimes (X^2)^T    \right) \\
    & + \gamma_w \Big[  e^{-s} \, \sigma_- \otimes (\sigma_+)^T \\
    &\qquad \qquad - \frac{1}{2} \left( \sigma_+ \sigma_- \otimes \mathds{1} + \mathds{1} \otimes (\sigma_+ \sigma_-)^T     \right)   \Big]  .
    \end{split}
\end{equation}
In the three-level case, derived from Eq.~\eqref{eq:lindblad_threelevel}, the structure is analogous
\begin{equation}
   \begin{split}
    \mathcal{L}_s =& -i \left( H \otimes \mathds{1} - \mathds{1} \otimes H^T  \right) \\
    &- \gamma_m \left(X^2 \otimes \mathds{1} - 2 X\otimes X^T + \mathds{1} \otimes (X^2)^T    \right) \\
    &+ \gamma_e \Big[ L_e \otimes (L_e^\dagger)^T \\
    &\qquad \quad - \frac{1}{2} \left( L_e^\dagger L_e \otimes \mathds{1} + \mathds{1} \otimes (L_e^\dagger L_e)^T     \right)   \Big] \\
    &+ \gamma_w \Big[ e^{-s} \, L_w \otimes (L_w^\dagger)^T  \\
    &\qquad \quad -\frac{1}{2} \left( L_w^\dagger L_w \otimes \mathds{1} + \mathds{1} \otimes (L_w ^\dagger L_w )^T      \right)   \Big] .
    \end{split}
\end{equation}

Additionally, the Hilbert-Schmidt inner product between two operators $\hat{A}$ and $\hat{B}$ corresponds, in Liouville space, to the standard vector inner product,
\begin{equation}
  \mathrm{tr}(A^\dagger B) = \langle\!\langle A | B \rangle\!\rangle .
\end{equation}
In particular, choosing $\hat{A} = \hat{\mathds{1}}$ yields
\begin{equation}
    \mathrm{tr}(\hat{B}) = \langle \! \langle \mathds{1} | B \rangle \! \rangle .
\end{equation}
We now apply this identity to the MGF,
\begin{equation}
    Z_t(s) = \mathrm{tr}[\hat{\rho}_s(t)] = \mathrm{tr} \left( e^{t \hat{\mathcal{L}}_s} [\hat{\rho}(0)] \right) .
\end{equation}
Identifying $\hat{B} = e^{t \hat{\mathcal{L}}_s } [\hat{\rho}_0]$, the MGF can be written in vectorized form as
\begin{equation}
     Z_t(s) = \langle \! \langle \mathds{1} | e^{t \mathcal{L}_s } | \rho(0) \rangle \! \rangle .
\end{equation}
Finally, using the spectral decomposition of $\mathcal{L}_s$, we obtain
\begin{equation}
     Z_t(s) = \sum_k e^{t \lambda_k(s)} \, \langle \! \langle \mathds{1} | r_k^{(s)} \rangle \! \rangle \, \langle \! \langle l_k^{(s)} | \rho(0) \rangle \! \rangle .
\end{equation}
This is the exact formula for the moment generating function from which the different regimes of interest we describe in the main text follow in appropriate limits.

\section{Perturbative Expansion in the Counting Field $s$}
\label{app:pert_expansion}

In this appendix we evaluate Eqs.~\eqref{eq:mean_with_correc} and~\eqref{eq:var_with_correc} in closed
form. We follow the recursive perturbative scheme for counting statistics~\cite{flindt_full_2004,BraggioKonig2006, FlindtNovotnyBraggio2008, FlindtNovotnyBraggio2010}, evaluated via the Drazin inverse (pseudoinverse) of the Liouvillian as developed in Ref.~\onlinecite{FlindtNovotnyBraggio2010} (see also Ref.~\onlinecite{LandiReview2024} for a pedagogical review in the quantum-jump setting).

We write the Taylor expansion of the tilted Liouvillian and of the
dominant eigenpair in powers of the counting field $s$ about $s=0$,
\begin{align}
  \mathcal{L}_s
    &= \mathcal{L}_0 + s\,\mathcal{L}_1 + s^2\,\mathcal{L}_2 + \mathcal{O}(s^3),
    \label{eq:app_L_expansion} \\
  \lambda_{\rm dom}(s)
    &= s\,\lambda_1 + s^2\,\lambda_2 + \mathcal{O}(s^3),
    \label{eq:app_lam_expansion} \\
  | r^{(s)}_{\rm dom} \rangle\!\rangle
    &= | r_0 \rangle\!\rangle + s\,| r_1 \rangle\!\rangle
       + s^2\,| r_2 \rangle\!\rangle + \mathcal{O}(s^3),
    \label{eq:app_r_expansion} \\
  \langle\!\langle l^{(s)}_{\rm dom} |
    &= \langle\!\langle l_0 | + s\,\langle\!\langle l_1 |
       + s^2\,\langle\!\langle l_2 | + \mathcal{O}(s^3),
    \label{eq:app_l_expansion}
\end{align}
where $\mathcal{L}_n = \frac{1}{n!}\,\partial_s^n \mathcal{L}_s \big|_{s=0}$, $| r_0 \rangle\!\rangle = | \rho_{\rm ss} \rangle\!\rangle$ and $\langle\!\langle l_0 | = \langle\!\langle \mathds{1} |$. Throughout we use the gauge fixed in the main text,
\begin{equation}
  \langle\!\langle \mathds{1} | r_{\rm dom}^{(s)} \rangle\!\rangle = 1
  \qquad \text{for all } s ,
  \label{eq:app_gauge}
\end{equation}
together with the biorthonormalization $\langle\!\langle l_{\rm dom}^{(s)} | r_{\rm dom}^{(s)} \rangle\!\rangle = 1$. The gauge~\eqref{eq:app_gauge} fixes the component of each $| r_n \rangle\!\rangle$ ($n\ge 1$) along the steady state,
\begin{equation}
  \langle\!\langle \mathds{1} | r_n \rangle\!\rangle = 0 ,
  \qquad n \ge 1 ,
  \label{eq:app_gauge_orders}
\end{equation}
i.e., the corrections to the (dominant) right eigenvector are trace-free. This is the key simplification afforded by this gauge choice.

\subsection{First order: mean}
 
Inserting the expansions~\eqref{eq:app_L_expansion}--\eqref{eq:app_r_expansion} into the eigenvalue equation
$\mathcal{L}_s | r^{(s)} \rangle\!\rangle = \lambda_{\rm dom}(s) | r^{(s)} \rangle\!\rangle$
and collecting the $\mathcal{O}(s)$ terms gives
\begin{equation}
  \mathcal{L}_0 | r_1 \rangle\!\rangle
  + \mathcal{L}_1 | \rho_{\rm ss} \rangle\!\rangle
  = \lambda_1 | \rho_{\rm ss} \rangle\!\rangle .
  \label{eq:app_order1}
\end{equation}
For the tilted Liouvillian of the main text only the counted jump term carries the counting field, so in the vectorized representation ${\mathcal{L}_1 = -\gamma_w \, L_w \otimes (L_w^\dagger)^T}$. Equation~\eqref{eq:app_order1} contains two unknowns, $\lambda_1$ and $| r_1 \rangle\!\rangle$, and we extract them in turn. Projecting with $\langle\!\langle \mathds{1} |$ and using $\langle\!\langle \mathds{1} | \mathcal{L}_0 = 0$ and $\langle\!\langle \mathds{1} | \rho_{\rm ss} \rangle\!\rangle = 1$ yields 
\begin{equation}
  \lambda_1
  = \langle\!\langle \mathds{1} | \mathcal{L}_1 | \rho_{\rm ss} \rangle\!\rangle
  = - \gamma_w \mathrm{tr}\!\left[ L_w^\dagger L_w \, \rho_{\rm ss} \right]
  \equiv - k_{\rm ss} ,
  \label{eq:app_lam1}
\end{equation}
with $k_{\rm ss}$ the stationary emission rate. This reproduces the leading-order LDT prediction for the mean rate, ${\langle N \rangle_t}_{\rm LDT} = -\lambda_1 = -\lambda_{\rm dom}'(0)$.

Rearranging Eq.~\eqref{eq:app_order1} gives a linear system for the first-order correction, $\mathcal{L}_0 | r_1 \rangle\!\rangle = -\big(\mathcal{L}_1 - \lambda_1\big) | \rho_{\rm ss} \rangle\!\rangle $. One would like to write $|r_1\rangle\!\rangle = -\mathcal{L}_0^{-1}(\mathcal{L}_1 - \lambda_1)|\rho_{\rm ss}\rangle\!\rangle$, but $\mathcal{L}_0$ is not invertible: it possesses the zero eigenvalue that we are perturbing around. A singular linear system $\mathcal{L}_0 | x \rangle\!\rangle = | b \rangle\!\rangle$ is, however, still solvable provided the right-hand side contains no component along the zero mode, i.e., provided $| b \rangle\!\rangle$ lies in the range of $\mathcal{L}_0$. Concretely, $| b \rangle\!\rangle$ must be annihilated by the left zero-eigenvector $\langle\!\langle \mathds{1} |$:
\begin{equation}
     \langle\!\langle \mathds{1} |\big(\mathcal{L}_1 - \lambda_1\big)|\rho_{\rm ss}\rangle\!\rangle = \langle\!\langle \mathds{1} | \mathcal{L}_1 | \rho_{\rm ss} \rangle\!\rangle - \lambda_1 = 0 ,
\end{equation}
which is satisfied automatically. 
 
To write the solution compactly we need an operator that inverts $\mathcal{L}_0$ on the part of the space where it is invertible (the complement of the zero mode) and does nothing on the zero mode itself. This is the \emph{Drazin} inverse $\mathcal{L}_0^{D}$. In the spectral decomposition $\mathcal{L}_0 = \sum_{k} \lambda_k |r_k\rangle\!\rangle\langle\!\langle l_k|$, it acts as
\begin{equation}
  \mathcal{L}_0^{D} = \sum_{k \neq 0} \frac{1}{\lambda_k}\, |r_k\rangle\!\rangle\langle\!\langle l_k| ,
  \label{eq:app_Drazin_Inverse}
\end{equation}
i.e., it ``inverts what can be inverted". Equivalently, $\mathcal{L}_0^{D}$ is defined by~\cite{LandiReview2024}
\begin{equation}
   \begin{split}
     \mathcal{L}_0^{D} | \rho_{\rm ss} \rangle\!\rangle &= 0 , \\ 
     \langle\!\langle \mathds{1} | \mathcal{L}_0^{D} &= 0 , \\
     \mathcal{L}_0 \mathcal{L}_0^{D} &= \mathcal{L}_0^{D} \mathcal{L}_0 = \mathcal{Q} ,
    \end{split}
\end{equation}
with $\mathcal{Q} = \mathds{1} - | \rho_{\rm ss} \rangle\!\rangle \langle\!\langle \mathds{1} |$ the projector onto the complement of the stationary mode. In practice $\mathcal{L}_0^{D}$  is obtained directly from
\begin{equation}
     \mathcal{L}_0^{D} = \big( \mathcal{L}_0 + \mathcal{P} \big)^{-1} - \mathcal{P} , \qquad \mathcal{P} = | \rho_{\rm ss} \rangle\!\rangle \langle\!\langle \mathds{1} | .
\end{equation}
Applying $\mathcal{L}_0^{D}$ to Eq.~\eqref{eq:app_order1} and using $\mathcal{L}_0^{D}\mathcal{L}_0 = \mathcal{Q}$ we get
\begin{equation}
     | r_1 \rangle\!\rangle = - \mathcal{L}_0^{D} \, \mathcal{L}_1 | \rho_{\rm ss} \rangle\!\rangle .
     \label{eq:app_r1}
\end{equation}
The analogous treatment of the left equation $\langle\!\langle l^{(s)} | \mathcal{L}_s = \lambda_{\rm dom}(s) \langle\!\langle l^{(s)} |$ at $\mathcal{O}(s)$ gives
\begin{equation}
  \langle\!\langle l_1 |
  = - \langle\!\langle \mathds{1} | \, \mathcal{L}_1 \, \mathcal{L}_0^{D} .
  \label{eq:app_l1}
\end{equation}
Finally, from $c_{\rm dom}(s) = \langle\!\langle l^{(s)} | \rho(0) \rangle\!\rangle$,  the initial-state term to the mean becomes
\begin{equation}
  - (\ln c_{\rm dom})'(0)
  = \langle\!\langle \mathds{1} | \, \mathcal{L}_1 \, \mathcal{L}_0^{D} | \rho(0) \rangle\!\rangle .
  \label{eq:app_mean_correction}
\end{equation}

Using $\mathcal{L}_0^{D} = \mathcal{L}_0^{D}\mathcal{Q}$, together with $\mathcal{Q} | \rho(0) \rangle\!\rangle = | \rho(0) - \rho_{\rm ss} \rangle\!\rangle$ and the integral representation of the Drazin inverse
\begin{equation}
  \mathcal{L}_0^{D} = - \int_0^\infty d\tau \; e^{\tau \mathcal{L}_0} \, \mathcal{Q} ,
  \label{eq:app_drazin_integral}
\end{equation}
the correction~\eqref{eq:app_mean_correction} can be written as
\begin{equation}
  - (\ln c_{\rm dom})'(0)
  = \int_0^\infty d\tau \, \big[ k(\tau) - k_{\rm ss} \big] ,
\end{equation}
with $k(\tau) = \langle\!\langle \mathds{1} | \mathcal{L}_1 | \rho (\tau) \rangle\!\rangle = \gamma_w \mathrm{tr}\!\left[ L_w^\dagger L_w \, \rho(\tau) \right]$, as quoted in the main text. 

\subsection{Second order: variance}
 Collecting now the $\mathcal{O}(s^2)$ terms of the right eigenvalue equation
gives
\begin{equation}
  \mathcal{L}_0 | r_2 \rangle\!\rangle
  = -\,( \mathcal{L}_1 - \lambda_1 ) | r_1 \rangle\!\rangle
    - ( \mathcal{L}_2 - \lambda_2 ) | \rho_{\rm ss} \rangle\!\rangle .
  \label{eq:app_order2}
\end{equation}
For the tilted Liouvillian of the main text, $\mathcal{L}_2 = -\frac{1}{2} \mathcal{L}_1$. Projecting with $\langle\!\langle \mathds{1} |$,
and using Eq.~\eqref{eq:app_gauge_orders},
\begin{equation}
  \begin{split}
  \lambda_2 &=
   \langle\!\langle \mathds{1} | \mathcal{L}_2 | \rho_{\rm ss} \rangle\!\rangle
    + \langle\!\langle \mathds{1} | \mathcal{L}_1 | r_1 \rangle\!\rangle \\
  &= \langle\!\langle \mathds{1} | \mathcal{L}_2 | \rho_{\rm ss} \rangle\!\rangle
    - \langle\!\langle \mathds{1} | \mathcal{L}_1 \mathcal{L}_0^{D} \mathcal{L}_1 | \rho_{\rm ss} \rangle\!\rangle .
    \end{split}
  \label{eq:app_lam2}
\end{equation}
The variance rate predicted by the leading-term in LDT is $\lambda''_{\rm dom}(0) = 2\lambda_2 \equiv v_{\rm ss}$. The first term is the bare (Poissonian) shot noise, the second the correction from temporal correlations of the emissions. The analogue of Eq.~\eqref{eq:app_mean_correction} for the variance is
\begin{equation}
  (\ln c_{\rm dom})''(0)
  = 2\,\langle\!\langle l_2 | \rho(0) \rangle\!\rangle
    - \langle\!\langle l_1 | \rho(0) \rangle\!\rangle^2 .
  \label{eq:app_lnc2}
\end{equation}
The second-order left eigenvector follows from collecting the $\mathcal{O}(s^2)$ terms of the left eigenvalue equation. Multiplying it from the right by $\mathcal{L}_0^{D}$ (which removes the $\lambda_2$ term) and inserting Eq.~\eqref{eq:app_l1} yields 
\begin{equation}
   \begin{split}
  \langle\!\langle l_2 | \mathcal{Q} \;
  = \; & \langle\!\langle \mathds{1} | \mathcal{L}_1 \mathcal{L}_0^{D} \mathcal{L}_1 \mathcal{L}_0^{D} \\
   & - \lambda_1 \langle\!\langle \mathds{1} | \mathcal{L}_1 (\mathcal{L}_0^{D})^2 \\ 
   & - \langle\!\langle \mathds{1} | \mathcal{L}_2 \mathcal{L}_0^{D} .
    \end{split}
  \label{eq:app_l2_Q}
\end{equation}
Unlike at first order, the component of $\langle\!\langle l_2 |$ along the stationary mode is not free: biorthonormality $\langle\!\langle l^{(s)} | r^{(s)} \rangle\!\rangle = 1$ at $\mathcal{O}(s^2)$, together with the gauge~\eqref{eq:app_gauge_orders}, fixes it to
\begin{equation}
  \langle\!\langle l_2 | \rho_{\rm ss} \rangle\!\rangle
  = - \langle\!\langle l_1 | r_1 \rangle\!\rangle
  = - \langle\!\langle \mathds{1} | \mathcal{L}_1 (\mathcal{L}_0^{D})^2 \mathcal{L}_1 | \rho_{\rm ss} \rangle\!\rangle .
  \label{eq:app_l2_kernel}
\end{equation}
Writing
$\langle\!\langle l_2 | = \langle\!\langle l_2 | \mathcal{Q}
+ \langle\!\langle l_2 | \rho_{\rm ss} \rangle\!\rangle \langle\!\langle \mathds{1} |$
and contracting with $| \rho(0) \rangle\!\rangle$ gives
$\langle\!\langle l_2 | \rho(0) \rangle\!\rangle$, and
Eq.~\eqref{eq:app_lnc2} becomes
\begin{equation}
  \begin{split}
  (\ln c_{\rm dom})''(0) \;
  = \; & 2\,\langle\!\langle \mathds{1} | \mathcal{L}_1 \mathcal{L}_0^{D} \mathcal{L}_1 \mathcal{L}_0^{D} | \rho(0) \rangle\!\rangle \\
     &- 2\lambda_1 \langle\!\langle \mathds{1} | \mathcal{L}_1 (\mathcal{L}_0^{D})^2 | \rho(0) \rangle\!\rangle \\
     & - 2\,\langle\!\langle \mathds{1} | \mathcal{L}_2 \mathcal{L}_0^{D} | \rho(0) \rangle\!\rangle \\
     &- 2\,\langle\!\langle \mathds{1} | \mathcal{L}_1 (\mathcal{L}_0^{D})^2 \mathcal{L}_1 | \rho_{\rm ss} \rangle\!\rangle \\
    & - \langle\!\langle \mathds{1} | \mathcal{L}_1 \mathcal{L}_0^{D} | \rho(0) \rangle\!\rangle^2 .
  \end{split}
  \label{eq:app_var_correction}
\end{equation}
The variance correction in Eq.~\eqref{eq:app_var_correction} splits into two physically distinct kinds of contributions. Terms ending in $\mathcal{L}_0^{D} | \rho(0) \rangle\!\rangle$ depend on the initial state and vanish when $\rho(0) = \rho_{\rm ss}$. They are the genuine initial-state memory. The fourth term is independent of the initial state and survives even when starting in the steady state. It is not an initial-state effect at all, it is a finite-time boundary effect. For the two-level system with dephasing operator $\hat{\sigma}_x$, this corresponds to Eq.~\eqref{eq:variance_sigma_x_ss}.

\subsection{Summary}
 Collecting the results, the long-time expansions of the mean and variance read
\begin{align}
  \langle N \rangle_t
    &\simeq
    k_{\rm ss}\, t
    + \langle\!\langle \mathds{1} | \mathcal{L}_1 \mathcal{L}_0^{D} | \rho(0) \rangle\!\rangle ,
    \label{eq:app_mean_summary}\\[2pt]
  \langle N^2 \rangle_t - \langle N \rangle_t^2
    &\simeq
    v_{\rm ss} \, t
    + (\ln c_{\rm dom})''(0) ,
    \label{eq:app_var_summary}
\end{align}
with $k_{\rm ss} = -\lambda_1$ from Eq.~\eqref{eq:app_lam1}, $v_{\rm ss} = 2\lambda_2$ from Eq.~\eqref{eq:app_lam2}, and $(\ln c_{\rm dom})''(0)$ from Eq.~\eqref{eq:app_var_correction}. All quantities are built from the steady state $\rho_{\rm ss}$ and the single Drazin inverse $\mathcal{L}_0^{D}$, requiring no diagonalization of the tilted Liouvillian at finite $s$. The same scaffolding extends to higher cumulants, each additional order introducing one further factor of $\mathcal{L}_0^{D}$ and, correspondingly, one higher moment of the stationary emission correlations~\cite{flindt_full_2004,BraggioKonig2006, FlindtNovotnyBraggio2008, FlindtNovotnyBraggio2010}.

\section{Beyond the Asymptotic Limit: Exact Finite-Time Transient Contributions}
\label{app:time_resolved}

The previous appendix determined the constants to which the transient contributions saturate. Here we derive their full time dependence: the exact, finite-time deviations of the mean and variance from their leading-order LDT growth,
\begin{align}
   \mathcal{O}_{\rm mean}(t) &\equiv \langle N \rangle_t - k_{\rm ss}\,t ,
  \label{eq:app_ft_defmean} \\
  \mathcal{O}_{\rm var}(t)  &\equiv \mathrm{Var}(N)_t - v_{\rm ss}\,t .
  \label{eq:app_ft_defvar}
\end{align}
Let $| \rho_s(t) \rangle\!\rangle = e^{t\mathcal{L}_s} | \rho(0) \rangle\!\rangle$ denote the tilted state, whose trace is the MGF, $Z_t(s) = \langle\!\langle \mathds{1} | \rho_s(t) \rangle\!\rangle$. Define the first two moment vectors
\begin{align}
  | n_1(t) \rangle\!\rangle &= - \partial_s | \rho_s(t) \rangle\!\rangle
  \big|_{s=0} , \\
  | n_2(t) \rangle\!\rangle &= \partial_s^2 | \rho_s(t) \rangle\!\rangle
  \big|_{s=0} ,
  \label{eq:app_ft_defn}
\end{align}
whose traces are the moments, $\langle N \rangle_t = \langle\!\langle \mathds{1} | n_1(t) \rangle\!\rangle$ and $\langle N^2 \rangle_t = \langle\!\langle \mathds{1} | n_2(t) \rangle\!\rangle$. Differentiating the evolution equation $\partial_t | \rho_s (t) \rangle\!\rangle = \mathcal{L}_s | \rho_s (t) \rangle\!\rangle$ once and twice with respect to $s$ at $s=0$ gives
\begin{align}
  \partial_t | n_1 \rangle\!\rangle
    &= \mathcal{L}_0 | n_1 \rangle\!\rangle
     - \mathcal{L}_1 | \rho(t) \rangle\!\rangle ,
    \label{eq:app_ft_n1}\\
  \partial_t | n_2 \rangle\!\rangle
    &= \mathcal{L}_0 | n_2 \rangle\!\rangle
     - 2\,\mathcal{L}_1 | n_1 \rangle\!\rangle
     + 2\,\mathcal{L}_2 | \rho(t) \rangle\!\rangle ,
    \label{eq:app_ft_n2}
\end{align}
with $ | n_1(0) \rangle\!\rangle = | n_2(0) \rangle\!\rangle = 0 $ and $| \rho(t) \rangle\!\rangle = e^{t\mathcal{L}_0} | \rho(0) \rangle\!\rangle$ the physical state. Additionally, the following two integral identities will prove useful. Since $\partial_\tau \big( \mathcal{L}_0^{D} e^{\tau\mathcal{L}_0} \big) = \mathcal{Q}\, e^{\tau\mathcal{L}_0} = e^{\tau\mathcal{L}_0}\mathcal{Q}$,
\begin{align}
  \int_0^t d\tau \; e^{\tau\mathcal{L}_0} \mathcal{Q}
    &= \mathcal{L}_0^{D} \big( e^{t\mathcal{L}_0} - \mathds{1} \big) ,
    \label{eq:app_ft_id1}\\
  \int_0^t d\tau \; \tau\, e^{\tau\mathcal{L}_0} \mathcal{Q}
    &= t\, \mathcal{L}_0^{D} e^{t\mathcal{L}_0}
     - \big(\mathcal{L}_0^{D}\big)^2 \big( e^{t\mathcal{L}_0} - \mathds{1} \big) ,
    \label{eq:app_ft_id2}
\end{align}
the second following from the first by integration by parts. As $t \to \infty$, Eq.~\eqref{eq:app_ft_id1} reduces to the integral representation~\eqref{eq:app_drazin_integral}.

\subsection{Mean}
Projecting Eq.~\eqref{eq:app_ft_n1} onto $\langle\!\langle \mathds{1} |$ and using $\langle\!\langle \mathds{1} | \mathcal{L}_0 = 0$ gives
\begin{equation}
  \frac{d \langle N \rangle_t}{dt}
  = - \langle\!\langle \mathds{1} | \mathcal{L}_1 | \rho(t) \rangle\!\rangle
  = k(t) ,
  \label{eq:app_ft_meanrate}
\end{equation}
so that, integrating,
\begin{equation}
    \langle N \rangle_t = -\int_0^t d\tau \,
    \langle\!\langle \mathds{1} | \mathcal{L}_1 e^{\tau \mathcal{L}_0}
    | \rho(0) \rangle\!\rangle .
\end{equation}
Decomposing the propagator as $e^{\tau \mathcal{L}_0} = \mathcal{P} + e^{\tau \mathcal{L}_0} \mathcal{Q}$ separates off the stationary rate,
\begin{equation}
    \langle N \rangle_t = k_{\rm ss}\, t -\int_0^t d\tau \,
    \langle\!\langle \mathds{1} | \mathcal{L}_1 e^{\tau \mathcal{L}_0}
    \mathcal{Q} | \rho(0) \rangle\!\rangle ,
\end{equation}
and applying the identity~\eqref{eq:app_ft_id1} yields the closed result
\begin{equation}
  \mathcal{O}_{\rm mean}(t)
  = \langle\!\langle l_1 | \rho(t) \rangle\!\rangle
  - \langle\!\langle l_1 | \rho(0) \rangle\!\rangle ,
  \label{eq:app_ft_meanresult}
\end{equation}
with $\langle\!\langle l_1 |$ from Eq.~\eqref{eq:app_l1}. As $t \to \infty$, $\rho(t) \to \rho_{\rm ss}$ and $\langle\!\langle l_1 | \rho_{\rm ss} \rangle\!\rangle = 0$, recovering the constant of Eq.~\eqref{eq:app_mean_summary}.

\subsection{Variance}

We now repeat the procedure at second order. Note first that the first moment vector can be decomposed into its component along the steady state and a traceless remainder,
\begin{equation}
  | n_1(t) \rangle\!\rangle
  = \langle N \rangle_t \, | \rho_{\rm ss} \rangle\!\rangle
  + | \tilde n_1(t) \rangle\!\rangle ,
  \;
  | \tilde n_1 \rangle\!\rangle = \mathcal{Q} | n_1 \rangle\!\rangle .
  \label{eq:app_ft_split}
\end{equation}
Applying $\mathcal{Q}$ to Eq.~\eqref{eq:app_ft_n1} and using $\mathcal{L}_0 \mathcal{P} = 0$, $\mathcal{Q}\mathcal{L}_0 = \mathcal{L}_0$,
\begin{equation}
  \partial_t | \tilde n_1 \rangle\!\rangle
  = \mathcal{L}_0 | \tilde n_1 \rangle\!\rangle
  - \mathcal{Q} \mathcal{L}_1 | \rho(t) \rangle\!\rangle ,
  \qquad | \tilde n_1(0) \rangle\!\rangle = 0 .
  \label{eq:app_ft_n1tilde}
\end{equation}
At long times, $\rho(t) \to \rho_{\rm ss}$, and Eq.~\eqref{eq:app_ft_n1tilde} relaxes to the fixed point $0 = \mathcal{L}_0 | \tilde n_1^\infty \rangle\!\rangle - \mathcal{Q}\mathcal{L}_1 | \rho_{\rm ss} \rangle\!\rangle$, i.e.,
\begin{equation}
  | \tilde n_1^\infty \rangle\!\rangle
  = \mathcal{L}_0^{D} \mathcal{L}_1 | \rho_{\rm ss} \rangle\!\rangle
  = - | r_1 \rangle\!\rangle ,
  \label{eq:app_ft_n1inf}
\end{equation}
by Eq.~\eqref{eq:app_r1}. Solving Eq.~\eqref{eq:app_ft_n1tilde} by variation of constants and separating $\rho(\tau) = \rho_{\rm ss} + \delta\rho(\tau)$ with $| \delta\rho(\tau) \rangle\!\rangle = e^{\tau\mathcal{L}_0}\mathcal{Q}| \rho(0) \rangle\!\rangle$ gives the exact solution
\begin{equation}
  | \tilde n_1(t) \rangle\!\rangle
  = - | r_1 \rangle\!\rangle + e^{t\mathcal{L}_0} | r_1 \rangle\!\rangle
  - | K(t) \rangle\!\rangle ,
  \label{eq:app_ft_n1sol}
\end{equation}
with $| K(t) \rangle\!\rangle  = \int_0^t d\tau\, e^{(t-\tau)\mathcal{L}_0} \mathcal{Q} \mathcal{L}_1 | \delta\rho(\tau) \rangle\!\rangle$. 

From Eqs.~\eqref{eq:app_ft_n1}--\eqref{eq:app_ft_n2} and the split~\eqref{eq:app_ft_split},
\begin{equation}
  \begin{split}
  \frac{d}{dt} \mathrm{Var}(N)_t
  =& \, 2 \langle\!\langle \mathds{1} | \mathcal{L}_2 | \rho(t) \rangle\!\rangle
  - 2 \langle\!\langle \mathds{1} | \mathcal{L}_1 | \tilde n_1(t) \rangle\!\rangle \\
  &- 2 \langle N \rangle_t \big[ k(t) - k_{\rm ss} \big] .
  \end{split}
  \label{eq:app_ft_varrate}
\end{equation}
In the stationary limit the last term vanishes and, using Eq.~\eqref{eq:app_ft_n1inf},
\begin{equation}
  \frac{d}{dt} \mathrm{Var}(N)_t
  \!\!\;\underset{t\to\infty}{\longrightarrow}\;\!\!
  2 \langle\!\langle \mathds{1} | \mathcal{L}_2 | \rho_{\rm ss} \rangle\!\rangle
  + 2 \langle\!\langle \mathds{1} | \mathcal{L}_1 | r_1 \rangle\!\rangle
  = v_{\rm ss} ,
\end{equation}
an independent re-derivation of Eq.~\eqref{eq:app_lam2}.

We integrate $d\mathrm{Var}(N)_t/dt - v_{\rm ss}$ term by term. The $\mathcal{L}_2$ term and the $e^{t\mathcal{L}_0}| r_1 \rangle\!\rangle$ part of Eq.~\eqref{eq:app_ft_n1sol} integrate with the identity~\eqref{eq:app_ft_id1} (note $\mathcal{Q}| r_1 \rangle\!\rangle = | r_1 \rangle\!\rangle$). For the $| K \rangle\!\rangle$ term, exchanging the order of the two time integrations and applying Eq.~\eqref{eq:app_ft_id1} to the inner one produces one closed contribution plus a remainder $X(t)$ defined below. In the cross term we write $\langle N \rangle_\tau = k_{\rm ss}\tau + \mathcal{O}_{\rm mean}(\tau)$ and $k(\tau) - k_{\rm ss} = \mathcal{O}_{\rm mean}'(\tau)$: the piece $\propto \tau\,\mathcal{O}_{\rm mean}'$ integrates with the identity~\eqref{eq:app_ft_id2}, while the remaining piece is a perfect derivative producing the term $- \mathcal{O}_{\rm mean}(t)^2$.

Finally, defining $\mathcal{F}(t) = e^{t\mathcal{L}_0} - \mathds{1}$, we get
\begin{equation}
  \begin{split}
  \mathcal{O}_{\rm var}(t)
  = \; & - 2\,\langle\!\langle l_2 | \mathcal{Q}\,
         \mathcal{F}(t) | \rho(0) \rangle\!\rangle \\
       & - 2\,\langle\!\langle \mathds{1} | \mathcal{L}_1 \mathcal{L}_0^{D}
         \mathcal{F}(t) | r_1 \rangle\!\rangle \\
       & + 2 k_{\rm ss}\, t \, \langle\!\langle \mathds{1} | \mathcal{L}_1
         \, \mathcal{L}_0^{D} e^{t\mathcal{L}_0} | \rho(0) \rangle\!\rangle \\
       & - \mathcal{O}_{\rm mean}(t)^2 + X(t) ,
  \end{split}
  \label{eq:app_ft_varresult}
\end{equation}
in terms of Eq.~\eqref{eq:app_l2_Q} and with 
\begin{equation}
  X(t) = 2 \int_0^t \!\! du \,
  \langle\!\langle \mathds{1} | \mathcal{L}_1 \mathcal{L}_0^{D}
  e^{(t-u)\mathcal{L}_0} \mathcal{L}_1 e^{u\mathcal{L}_0} \mathcal{Q}
  | \rho(0) \rangle\!\rangle .
  \label{eq:app_ft_X}
\end{equation}
Every term of Eq.~\eqref{eq:app_ft_varresult} vanishes at $t=0$, as it must, and $X(0) = X(\infty) = 0$. In the limit $t \to \infty$ one has $\mathcal{F}(t) \to -\mathcal{Q}$, $t \mathcal{L}_0^{D} e^{t\mathcal{L}_0} \to 0$, and $\mathcal{O}_{\rm mean}(t) \to -\langle\!\langle l_1 | \rho(0) \rangle\!\rangle$, and Eq.~\eqref{eq:app_ft_varresult} reproduces, term by term, the five contributions of Eq.~\eqref{eq:app_var_correction}.

\subsection{Two-level system with $\hat X = \hat\sigma_x$}

For the two-level model with $\hat X(\theta=0) = \hat\sigma_x$ the free relaxation is governed by a single rate $\Gamma \equiv 4\gamma_m + \gamma_w$. Evaluating Eq.~\eqref{eq:app_ft_meanresult} for the ground and excited initial states
\begin{equation}
  \mathcal{O}_{\rm mean}(t)
  = \Delta_{\rm mean} \big( 1 - e^{-\Gamma t} \big) ,
  \label{eq:app_ft_mean2lvl}
\end{equation}
with the asymptotic constants $\Delta_{\rm mean}^{(g)} = - 2\gamma_m\gamma_w / \Gamma^2$ and $\Delta_{\rm mean}^{(e)} = \gamma_w (2\gamma_m + \gamma_w)/\Gamma^2$ of Eq.~\eqref{eq:mean_sigma_x_ground_early} and Eq.~\eqref{eq:mean_sigma_x_excited_early}, respectively. Thus, $\mathcal{O}_{\rm mean}(\infty)= \Delta_{\rm mean}\equiv\langle N \rangle_{\rm early}$.

Similarly, the variance in Eq.~\eqref{eq:app_ft_varresult} reduces to
\begin{equation}
  \mathcal{O}_{\rm var}(t)
  = c
  + \big[ \Delta_{\rm mean}^2 - c + b\,t \big] e^{-\Gamma t}
  - \Delta_{\rm mean}^2 \, e^{-2\Gamma t} ,
  \label{eq:app_ft_var2lvl}
\end{equation}
where $c$ denotes the asymptotic value given in Eqs.~\eqref{eq:variance_sigma_x_ground_early} and~\eqref{eq:variance_sigma_x_excited_early}, and the coefficients are $b^{(g)} = -16\gamma_m^2\gamma_w^2/\Gamma^3$ and $b^{(e)} = 8\gamma_m\gamma_w^2(2\gamma_m+\gamma_w)/\Gamma^3$, for the ground and excited states, respectively. For the steady-state preparation, $\mathcal{O}_{\rm mean}^{(\rm ss)}(t) = 0$ while
\begin{equation}
  \mathcal{O}_{\rm var}^{(\rm ss)}(t)
  = c^{(\rm ss)} \big( 1 - e^{-\Gamma t} \big) ,
\end{equation}
with $c^{(\rm ss)}$ given by Eq.~\eqref{eq:variance_sigma_x_ss}. Equations~\eqref{eq:app_ft_mean2lvl} and~\eqref{eq:app_ft_var2lvl} are the analytical curves shown in Fig.~\ref{fig:2level_evolution_mean_variance}.

\section{Beyond Autonomous: Extension to Driven Emitters\label{app:driven}}

In this appendix, we address what role the initial-state transient terms play in the precision with which a general parameter $\varepsilon$ can be estimated. We compute the Fisher information associated with the photon count and identify where the transient terms enter. We find that they do contribute to the Fisher information, but only at the same order as the third and higher cumulants.

Consider a two-level system with transition frequency $\Omega$, subject to a static longitudinal field and a transverse drive of amplitude $\gamma_d$ and frequency $\omega_d$,
\begin{equation}
  \hat{H}_d(t)=  \Omega \ket{e}\bra{e} +  \gamma_d \sin \theta \hat{\sigma}_z +2 \gamma_d \cos\theta \cos(\omega_d t) \hat{\sigma}_x .
\end{equation}
Moving to the rotating frame via $\hat U(t) = e^{i\omega_d t\ket{e}\bra{e}}$ and discarding counter-rotating terms, we get 
\begin{equation}
  \hat{H}_d^{\mathrm{RWA}} =  \Delta \ket{e}\bra{e} +  \gamma_d \hat{X}(\theta) ,
\end{equation}
with $\Delta = \Omega- \omega_d$ and $\hat{X}(\theta)$ as in Eq.~\eqref{eq:X_2level}. The approximation is valid for $\gamma_d, |\Delta|, \gamma_w \ll \omega_d$. Using $\hat\sigma_z = 2\ket{e}\bra{e} - \hat{\mathds 1}$, we obtain, up to an irrelevant identity term,
\begin{equation}
    \hat{H}_d^{\mathrm{RWA}} = \Lambda(\theta) \ket{e}\bra{e} + g(\theta) \hat{\sigma}_x ,
\end{equation}
with an effective splitting $\Lambda(\theta) = \Delta + 2\gamma_d\sin\theta$ and $g(\theta) = \gamma_d\cos\theta$. The tilted Lindbladian for this driven setup, including the fluorescence measurement, is given by,
\begin{equation}
    \begin{split}
    \hat{\mathcal{L}}_s[\hat{\rho}] &= -i [\hat{H}_d^{\mathrm{RWA}}, \hat{\rho}] -\gamma_m [\hat{X}, [\hat{X},\hat{\rho}]] \\
    &+ \gamma_w \Big[  e^{-s} \, \hat{\sigma}_- \hat{\rho} \, \hat{\sigma}_+ - \frac{1}{2} \left( \hat{\rho} \, \hat{\sigma}_+  \hat{\sigma}_- +  \hat{\sigma}_+ \hat{\sigma}_- \, \hat{\rho}  \right) \Big] .
   \end{split}
\end{equation}
The dephasing induced by $\hat{X}(\theta)$ is understood to be defined in the rotating frame. The emission channel requires no such convention: the frame transformation acts as $\hat{\sigma}_-\to e^{-i\omega_d t}\hat{\sigma}_-$, so the phase factors cancel in the jump term $\hat{\sigma}_-\hat{\rho}\,\hat{\sigma}_+$ while $\hat{\sigma}_+\hat{\sigma}_-$ is unchanged. The counting field $s$ therefore enters the generator identically in both frames, and the photon-counting statistics are frame-independent. In what follows, we omit the RWA subscript on the Hamiltonian.

Introducing the quantities $\zeta = \frac{1}{2}\gamma_w + 4\gamma_m$ and $\alpha(\theta) \equiv  \Delta\cos\theta$, the steady state again takes the form of Eq.~\eqref{eq:2level_density_ss}, now with 
\begin{align}
    p_e^{\rm ss} &= \frac{4\gamma_w\big(g(\theta)^2 + \gamma_m \zeta \cos^2\theta\big)
                          + 8\gamma_m \alpha(\theta)^2}{\mathcal{N}(\theta)} ,
                          \label{eq:app_driven_pe}\\[1ex]
    \rho_u &= -\frac{4\gamma_w\big(g(\theta)\Lambda(\theta) + \gamma_m\zeta \sin 2\theta\big)}{\mathcal{N}(\theta)} , \\[1ex]
    \rho_v &= \frac{2\gamma_w\big(g(\theta)\gamma_w - 4\gamma_m \alpha(\theta) \sin\theta\big)}{\mathcal{N}(\theta)} ,
\end{align}
where the common denominator reads
\begin{equation}
    \begin{split}
    \mathcal{N}(\theta) = 
    & 4\gamma_w\big(\Lambda(\theta)^2 + 2g(\theta)^2\big) + 16\gamma_m\alpha(\theta)^2 \\
    &\quad +2\gamma_w\zeta \Big[\gamma_w + 4\gamma_m\big(1 + \sin^2\theta\big)\Big] .
    \end{split}
\end{equation}

\begin{figure}[htbp]
   \includegraphics[width=0.45\textwidth]{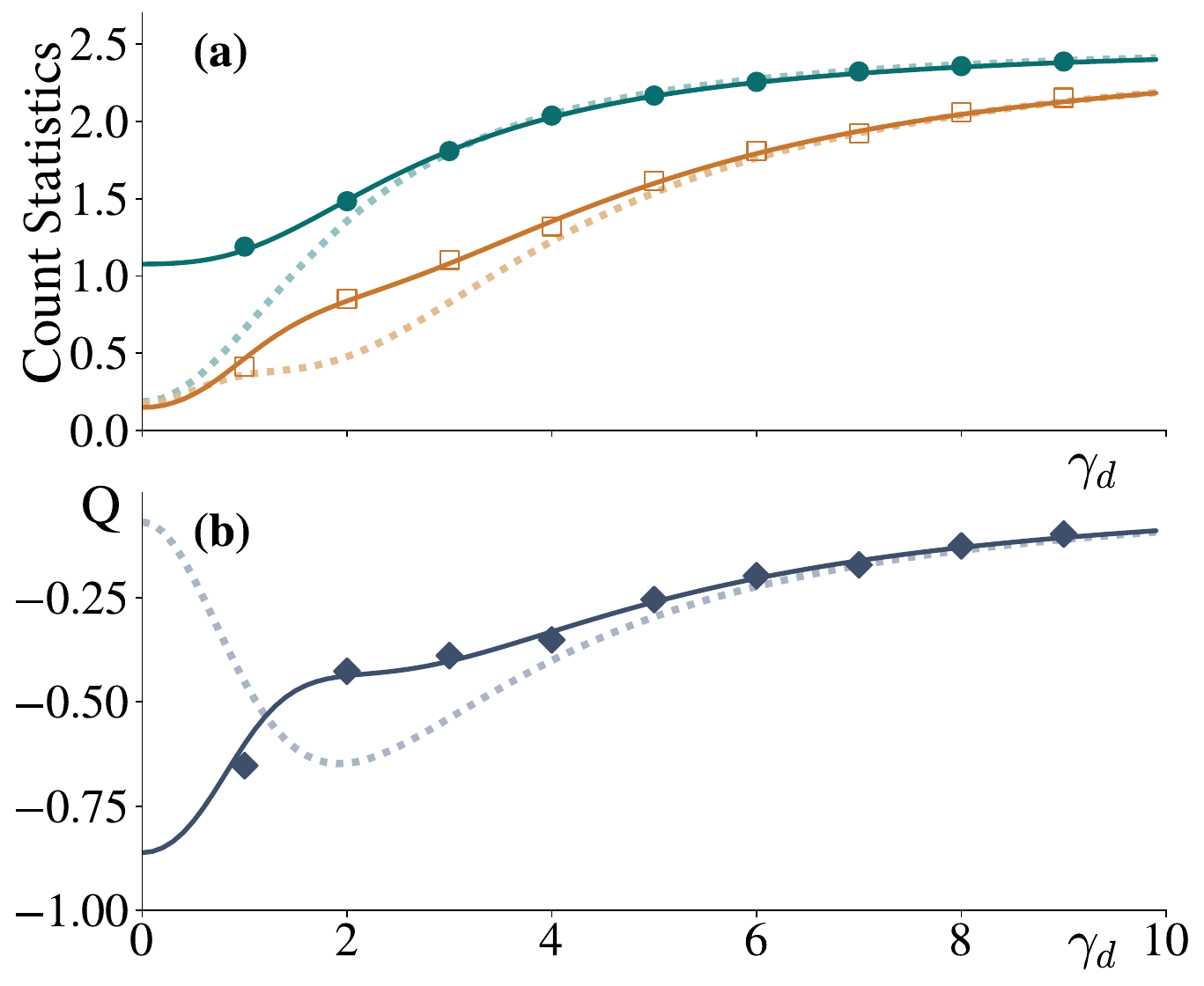}
   \caption{
    (a) Mean number (green) and variance (orange) of clicks as functions of the drive amplitude $\gamma_d$ for the dephasing operator $\hat{X}(\theta=0) = \hat{\sigma}_x$. 
    (b) Mandel $Q_t$ parameter as a function of $\gamma_d$ (dark blue).
    In both panels, dashed lines denote the LDT predictions, while solid lines include both the LDT and transient contributions arising from the initial state (Eqs.~\eqref{eq:app_mean_summary} and~\eqref{eq:app_var_summary} evaluated analytically). Scatter points correspond to numerical simulations performed using the Kraus-operator formalism (see Appendix~\ref{app:Kraus_Formalism}). Parameters: $\Delta=0$, $\gamma_w = 5$, $\gamma_m=0.1$. The simulations are performed with time step $dt = 0.001$ over $1000$ steps using $10^6$ trajectories. Initial state $\hat{\rho}(0) = \ket{e}\bra{e}$. }
   \label{fig:2level_driven_mean_variance_excited_gd}
\end{figure}

For concreteness, we set $\Delta=\theta=0$, in which case the Hamiltonian reduces to $\hat H_d=\gamma_d\hat\sigma_x$. For this model, the long-time emission and variance rates, ${\langle N \rangle_t}_{\mathrm{LDT}}=k_{\rm ss}t$ and $\mathrm{Var}_{\rm LDT}(N)_t=v_{\rm ss}t$ can be obtained in closed form. Defining $D=8\gamma_d^2+\left(4\gamma_m+\gamma_w\right)\left(8\gamma_m+\gamma_w\right)$, we find
\begin{align}
  k_{\rm ss} &= \frac{2\gamma_w}{D} \left(2\gamma_d^2+\gamma_m\left(8\gamma_m+\gamma_w\right)\right), \ \label{eq:app_kss}\\[4pt]
  v_{\rm ss} &= \frac{k_{\rm ss}}{D^2}
    \Big[64\gamma_d^4 \nonumber + 8\gamma_d^2\left(64\gamma_m^2+16\gamma_m\gamma_w-\gamma_w^2\right) \nonumber\\
  &\quad + \left(8\gamma_m+\gamma_w\right)^2
      \left(16\gamma_m^2+4\gamma_m\gamma_w+\gamma_w^2\right)\Big],
  \label{eq:app_vss}
\end{align}
where $k_{\rm ss} = \gamma_w p_e^{\rm ss}$ can be read directly from Eq.~\eqref{eq:app_driven_pe} while $v_{\rm ss}$ follows from Eq.~\eqref{eq:app_lam2}. These expressions, together with the transient contributions, are shown in Fig.~\ref{fig:2level_driven_mean_variance_excited_gd}.

With the emission and variance rates in hand, we now turn to the precision with which a general parameter $\varepsilon$ can be estimated from the count record. Inverting Eq.~\eqref{eq:MGF_trace}, the probability of observing $N$ jumps, given in Eq.~\eqref{eq:Pt(N)}, can be written as~\cite{LandiReview2024}:
\begin{equation}
  \begin{split}
  \mathcal{P}_t(N,\varepsilon)
  &= \frac{1}{2\pi i}\int_{-i\pi}^{i\pi} ds\; e^{sN} Z_t(s,\varepsilon) \\
  &= \frac{1}{2\pi i}\int_{-i\pi}^{i\pi} ds\; e^{\Phi(s,\varepsilon)},
  \end{split}
\end{equation}
where, in the second line, we have used Eq.~\eqref{eq:Zt(s)} and $\Phi(s,\varepsilon)$ is given by
\begin{equation}
   \Phi(s,\varepsilon) = t \left[ s\, n + \lambda_{\rm dom}(s,\varepsilon) \right] + \ln c_{\rm dom}(s,\varepsilon), 
\end{equation}
with $N=nt$. The bracketed term is of order $t$, whereas $\ln c_{\rm dom}$, which carries the entire initial-state dependence, is of order unity. Because the extensive part dominates, the integrand is sharply peaked around the saddle point $s^\ast$~\cite{Daniels1954,LandiReview2024}, determined by the stationarity condition to the leading order,
\begin{equation}
    n + \lambda_{\rm dom}'(s^\ast) =0 .
    \label{eq:app_stat_condition}
\end{equation}
At the saddle-point, $t \left[ s^\ast \, n + \lambda_{\rm dom}(s^\ast) \right]=-t I(n) $, where $I(n)$ is the \emph{large deviation function}~\footnote{
The large deviation function $I(n)$ characterizes the exponential decay of the probability $\mathcal{P}_t(N)\asymp e^{-tI(n)}$ as $t\to\infty$. More generally, the connection between $I(n)$ and the scaled cumulant generating function (SCGF), $\theta(s)=\lim_{t\to\infty}t^{-1}\ln Z_t(s)$, which in the present setting coincides with the dominant eigenvalue $\lambda_{\rm dom}(s)$, is established by the \textit{G\"{a}rtner-Ellis theorem}~\cite{Touchette2009,Touchette2018}: when the SCGF exists and satisfies the appropriate differentiability conditions, $I(n)$ is given by the Legendre-Fenchel transform $I(n)=\sup_s\left[-sn-\theta(s)\right]$.
}.

Expanding to second order about $s^\ast$, 
\begin{equation}
   \begin{split}
    \Phi(s,\varepsilon) \simeq
     &-t\,I(n,\varepsilon) + \ln c_{\rm dom}(s^\ast)  \\
     &+ \frac{t}{2}\lambda_{\rm dom}''(s^\ast,\varepsilon)\,(s-s^\ast)^2 .
  \end{split}
\end{equation}
Using the $2\pi i$-periodicity of the integrand, we shift the original contour to the vertical line $\mathrm{Re}\,s = s^\ast$. Parameterizing
$s=s^\ast+iy$ with $y$ real gives the quadratic term $-\tfrac{t}{2}\lambda_{\rm dom}''(s^\ast)y^2$, so that the probability then reads as
\begin{equation}
  \begin{split}
  &\mathcal{P}_t(N,\varepsilon)  \\
  &\simeq  \frac{e^{-tI(n,\varepsilon)}\,c_{\rm dom}(s^\ast,\varepsilon)}{2\pi}
   \int_{-\infty}^{\infty} \! dy\;e^{-\frac{t}{2}\lambda_{\rm dom}''(s^\ast,\varepsilon)y^2} \\[1ex]
  &= \frac{e^{-tI(n,\varepsilon)}\,c_{\rm dom}(s^\ast,\varepsilon)}
  {\sqrt{2\pi t\,\lambda_{\rm dom}''(s^\ast,\varepsilon)}} ,
  \end{split}
\end{equation}
where the integration has been extended to the whole real line, which is exponentially accurate for $t\,\lambda_{\rm dom}''(s^\ast)\gg1$. Note that the above result using the saddle-point method is only an approximation; in particular, the probability distribution need not be normalized in the saddle point approximation~\cite{LandiReview2024}. Nevertheless, it is a good approximation to obtain leading order estimates which we are interested in. Taking the logarithm, we have,
\begin{equation}
   \begin{split}
  \ln \mathcal{P}_t(N, \varepsilon)  =
  &-t\,I(n, \varepsilon) + \ln c_{\rm dom}(s^\ast, \varepsilon) \\
  &- \frac{1}{2} \ln\!\big[2\pi t\lambda_{\rm dom}''(s^\ast, \varepsilon)\big] + \dots
  \end{split}
\end{equation}
Each term of this expansion depends on $\varepsilon$ both explicitly and implicitly, through the saddle location $s^\ast(\varepsilon)$. The derivative $\partial_\varepsilon\ln\mathcal{P}_t(N,\varepsilon)$, taken at fixed $N$, measures how much more or less likely the observed count becomes when $\varepsilon$ shifts. In differentiating the leading term the implicit dependence drops out by the stationarity condition in Eq.~\eqref{eq:app_stat_condition}, and expanding near $s=0$, we get that
\begin{equation}
    \begin{split}
    \partial_\varepsilon I(n, \varepsilon) 
    &= -\partial_\varepsilon \lambda_{\rm dom}(s^\ast, \varepsilon) \simeq - s^\ast\,\partial_s \partial_\varepsilon \lambda_{\rm dom}(0,\varepsilon) \\
    &= -s^\ast\,\partial_\varepsilon\big[\lambda_{\rm dom}'(0,\varepsilon)\big]
     = \,s^\ast\, \partial_\varepsilon k_{\rm ss} .
    \end{split}
\end{equation}
Inverting the saddle condition near $s=0$ in Eq.~\eqref{eq:app_stat_condition} gives $s^\ast \simeq -(n-k_{\rm ss})/v_{\rm ss}$ (the next order involves the third cumulant rate $\lambda_{\rm dom}'''$), so that 
\begin{equation}
   \begin{split}
  \partial_\varepsilon \ln \mathcal{P}_t(N, \varepsilon)  \,=\,
  &\frac{\partial_\varepsilon k_{\rm ss}}{v_{\rm ss}}\big(N-k_{\rm ss}t\big) \\
  &+ \partial_\varepsilon \ln c_{\rm dom}(s^\ast, \varepsilon) \\
  &- \frac{1}{2} \partial_\varepsilon \ln\!\big[2\pi t\lambda_{\rm dom}''(s^\ast, \varepsilon)\big] + \dots
  \end{split}
\end{equation}
The three terms scale differently in $t$. The first is $\mathcal{O}(\sqrt t)$,
since $N-k_{\rm ss}t\approx \sqrt{v_{\rm ss}t}$; combined with the saddle condition
this also fixes $s^\ast\approx -(N-k_{\rm ss}t)/(v_{\rm ss}t) \sim \mathcal{O}(t^{-1/2})$. The second is a full power of $t$ smaller: since normalization imposes $c_{\rm dom}(0,\varepsilon)=1$, the initial-state term goes as $\partial_\varepsilon \ln c_{\rm dom}(s^\ast,\varepsilon)
\approx
s^\ast \,\partial_s\partial_\varepsilon
\ln c_{\rm dom}(0,\varepsilon)
\sim
\mathcal{O}\!\left(t^{-1/2}\right)$. Finally, the third term is time-independent because $\partial_\varepsilon
\ln [t \lambda_{\rm dom}''(s^\ast,\varepsilon)]
\approx \partial_\varepsilon
\ln v_{\rm ss}(\varepsilon).
$
Therefore, upon squaring and averaging, these terms contribute to the Fisher information, $\text{F}(\varepsilon) = \langle[\partial_\varepsilon \ln \mathcal{P}_t(N, \varepsilon)]^2\rangle$ yielding
\begin{equation}
  \text{F}(\varepsilon) = \frac{(\partial_\varepsilon k_{\rm ss})^2}{v_{\rm ss}} t + \mathcal{C} + \mathcal{O}(t^{-1}) .
  \label{eq:fisher_rate}
\end{equation}
The transient constant terms of Eqs.~\eqref{eq:mean_with_correc} and \eqref{eq:var_with_correc} do appear at the time-independent term $\mathcal{C}$, but there they sit alongside every other subleading contribution: the term $\partial_\varepsilon\ln v_{\rm ss}$ and the third and higher cumulants. As anticipated above, isolating the time-independent $\mathcal{C}$ term of $\text{F}(\varepsilon)$ would require the full hierarchy.

The leading rate nonetheless dominates at long times, so it is meaningful to ask how much of the available information it captures. We can benchmark Eq.~\eqref{eq:fisher_rate} against the \textit{quantum} Fisher information (QFI), the maximum Fisher information attainable by any measurement on the system and its output fields. Consider the case where we want to estimate the drive amplitude, with $\varepsilon=\gamma_d$. For a parameter appearing only in the Hamiltonian, the QFI over a long time $t$ is~\cite{Gammelmark2014,LandiReview2024}
\begin{equation}
  \text{QFI}(\gamma_d) = -4 \, t\,
  \mathrm{tr}\!\Big[(\partial_{\gamma_d} \hat H_d)\,\hat{\mathcal{L}}_0^{D}
  \big\{\hat{\rho}_{\rm ss},\,\partial_{\gamma_d} \hat H_d\big\}\Big] ,
\end{equation}
where $\hat{\mathcal{L}}_0^{D}$ is the Drazin inverse introduced in Eq.~\eqref{eq:app_Drazin_Inverse}. With $\partial_{\gamma_d}\hat H=\hat\sigma_x$ it gives 
\begin{equation}
  \text{QFI}(\gamma_d)=\frac{16}{\gamma_w} t,
  \label{eq:app_QFI}
\end{equation}
independent of $\gamma_d$ and $\gamma_m$.

Consider first $\gamma_m=0$. Dividing the leading term of the Fisher information in Eq.~\eqref{eq:fisher_rate} by the QFI~\eqref{eq:app_QFI},
\begin{equation}
 \frac{
 \text{F}(\gamma_d)
 }{\text{QFI}(\gamma_d)} = \frac{\gamma_w^6}
  {512\gamma_d^6+\gamma_w^6}
  \;\xrightarrow[\ \gamma_d\to0\ ]{}\; 1 ,
\end{equation}
implying that in the weak-drive limit counting the photons extracts all the information the field carries about $\gamma_d$. If the two-level system is driven too hard, $p_e^{\rm ss}$ saturates to $1/2$ and the count rate stops responding to changes in $\gamma_d$.

In the general case, $\gamma_m\neq 0$, the ratio is more involved, but for
$\gamma_m\ll\gamma_w$ it factorizes as
\begin{equation}
  \frac{\mathrm{F}(\gamma_d)}{\mathrm{QFI}(\gamma_d)}
  \simeq
  \frac{2\gamma_d^{2}}{2\gamma_d^{2}+\gamma_m\gamma_w}
  \frac{\gamma_w^{6}}{512\gamma_d^{6}+\gamma_w^{6}} ,
\end{equation}
the second factor being the exact result at $\gamma_m=0$ and the first the additional suppression from dephasing. Since dephasing also pumps the two-level system, photons are emitted even at $\gamma_d = 0$. These carry no information on $\gamma_d$, yet still contribute shot noise. The two factors act in opposite directions: the first requires driving hard enough that $\gamma_d^{2}\gg\gamma_m\gamma_w/2$, so that drive-induced emission dominates the dephasing-pumped background, while the second requires $\gamma_d^{2}\ll\gamma_w^{2}/8$ to avoid saturation. Counting is therefore near-optimal only in the window $\gamma_m\gamma_w/2 \ll \gamma_d^{2} \ll \gamma_w^{2}/8$, which requires $\gamma_m \ll \gamma_w$.

\bibliography{references}
\end{document}